\documentclass{emulateapj}

\newcommand{\bov}{\mbox{\boldmath $v$}}
\newcommand{\bw}{\mbox{\boldmath $w$}}
\newcommand{\bnabla}{\mbox{\boldmath $\nabla$}}

\shorttitle{Interstellar Turbulent Pressure}
\shortauthors{Joung, Mac Low, \& Bryan}

\begin{document}

\title{Dependence of Interstellar Turbulent Pressure on Supernova Rate}

\author{M.~Ryan~Joung\altaffilmark{1}, Mordecai-Mark~Mac~Low\altaffilmark{2,3}, and Greg~L.~Bryan\altaffilmark{3}}
\altaffiltext{1}{Department of Astrophysical Sciences, Princeton University, Peyton Hall, Ivy Lane, NJ~08544; joung@astro.princeton.edu}
\altaffiltext{2}{Department of Astrophysics, American Museum of Natural History, 79th Street at Central Park West, New York, NY~10024; mordecai@amnh.org}
\altaffiltext{3}{Department of Astronomy, Columbia University, 550 West 120th Street, New York, NY~10027; gbryan@astro.columbia.edu}

\begin{abstract}
Feedback from massive stars is one of the least understood aspects of 
galaxy formation. 
   We
perform a suite of vertically stratified local interstellar medium
(ISM) models in which supernova rates and vertical gas column  
densities are systematically varied based on the Schmidt-Kennicutt law. 
Our simulations have a sufficiently high spatial resolution (1.95 pc) 
to follow the hydrodynamic interactions among multiple supernovae 
that structure the interstellar medium. 
At a given supernova rate, we find that the mean 
mass-weighted sound speed and velocity dispersion decrease as 
the inverse square root of gas density. 
The sum of  
thermal and turbulent pressures is nearly 
constant in the midplane, 
    so
the effective equation of state is isobaric. 
In contrast, across our four models having supernova rates that 
range from 
   one 
to 512 times the Galactic supernova rate, 
the mass-weighted velocity dispersion 
   remains in the range 4--6~km~s$^{-1}$.
Hence, gas averaged over $\sim$100 pc regions follows $P \propto
\rho^{\, \alpha}$ with $\alpha \approx 1$, 
indicating that the effective equation of state on this scale is 
close to isothermal. 
    Simulated 
\ion{H}{1} emission lines have widths of 10--18 km s$^{-1}$, comparable 
to observed values.  
    In our highest supernova rate model, superbubble blow-outs occur, and 
    the turbulent pressure on large scales is $\gtrsim$4 times higher than 
    the thermal pressure.
We find a tight correlation between the thermal and turbulent pressures 
averaged over $\sim$100 pc regions in the midplane of each model, 
as well as across the four ISM models.  
We construct a subgrid model for turbulent pressure based on 
analytic arguments and explicitly calibrate it against 
our stratified ISM simulations.  The subgrid model provides 
a simple yet physically motivated way to include 
supernova feedback in cosmological simulations.
\end{abstract}


\section{Introduction}

The observed vertical distribution of gas in the Galaxy, especially the cold 
neutral \ion{H}{1} component at large heights, cannot be supported by thermal pressure 
alone (Lockman \& Gehman 1991; Ferrara 1993). 
Often invoked to provide the required additional pressure are turbulent 
motions of the interstellar gas, magnetic fields, and cosmic rays (Boulares 
\& Cox 1990; Lockman \& Gehman 1991). Of these, turbulent pressure is 
perhaps the dominant pressure component in the interstellar medium (ISM; McKee 1990). 
There are strong reasons to believe that supersonic turbulence also regulates 
star formation processes in molecular clouds 
by lowering the efficiency of star formation 
(Elmegreen \& Scalo 2004; Mac Low \& Klessen 2004; 
Krumholz \& McKee 2005; Krumholz \& Tan 2007).

Supernova (SN) explosions are thought to be the dominant agent for 
producing turbulent motions in star-forming parts of galaxies 
(Norman \& Ferrara 1996; Mac Low \& 
Klessen 2004). Interactions 
between blast waves from multiple SNe lead to compression and shearing 
of the gas. In extreme cases, explosions that are spatially and temporally 
correlated may collectively generate a 
superwind, observed at both low (e.g., Martin 1999; Heckman et al.\ 2000) 
and high redshifts 
(Pettini et al.\ 2001, 2002; Shapley et al.\ 2003; Weiner et al.\ 2008). 
McKee \& Ostriker (1977) provided a 
picture of the ISM in which three distinct 
phases regulated by SN explosions exist in rough pressure equilibrium. 
The multiphase model 
does not directly include turbulent pressure 
produced by interactions among multiple SN-driven blast waves 
and large pressure fluctuations in the ISM.   
Numerical simulations 
(Rosen \& Bregman 1995; Korpi et al.\ 1999; 
Gazol-Pati\~no \& Passot 1999; de Avillez \& Berry 2001; 
Kim 2004; de Avillez \& Breitschwerdt 2004, 2005; Mac Low et al.\ 2005; 
Slyz et al.\ 2005; 
Joung \& Mac Low 2006, hereafter Paper I; Tasker \& Bryan 2006) 
and observations of 
thermally unstable gas (Heiles 2001; Heiles \& Troland 2003, 2005) also support this 
dynamic picture. Since 
turbulent pressure is a significant, sometimes the dominant, form of 
pressure in the ISM, its inclusion is vital for any realistic model of the ISM. 

Another motivation for this study comes from longstanding problems in cosmology.
The standard cold dark matter paradigm has been widely successful in explaining 
structures on cosmological scales (Cen et al.\ 1994; Zhang et al.\ 1995; Rauch 
1998; Bertschinger 1998; Cen \& Ostriker 2000; 
Springel et al.\ 2005; Dunkley et al.\ 2008).  
Yet our understanding of 
the luminous matter 
on galactic 
scales 
remains seriously incomplete, mainly due to complexities of 
baryon physics, for which 
additional nonlinear thermal and hydrodynamic processes must be taken 
into account. 
A prime example is the galaxy luminosity function, 
which shows lower abundances of galaxies 
both on the faint and the bright ends 
compared to the halo mass distribution 
(e.g., Benson et al.\ 2003).  
Other outstanding astrophysical problems 
include incorrect predictions for the 
size of disk galaxies 
(the ``angular momemtum problem''; e.g., Navarro \& White 1994; 
Steinmetz \& Navarro 1999) and for the fraction of gas 
that should have cooled to form galaxies (the ``overcooling problem''). 

Feedback from massive stars and active galactic nuclei 
plays a central role to explain these discrepancies 
in recent semi-analytic models 
(Croton et al.\ 2006; Bower et al.\ 2006; Hopkins et al.\ 2008; Somerville et al.\ 2008). 
Despite mounting evidence that SNe played a major role in 
galaxy formation (see, e.g., Adelberger et al.\ 2003 and 
references therein), it has been 
difficult to incorporate the details of stellar feedback into cosmological 
simulations. Early studies found that thermal energy input did not affect 
the ISM significantly, as it was radiated away almost instantly (Katz 1992; 
Navarro \& White 1993).  
This was due to a lack of
numerical resolution, which led to an absence of hot gas that would have 
radiated inefficiently and substantially decreased the overall cooling rate. 
Past attempts to resolve this difficulty (e.g., Kauffman et al.\ 
1993; Somerville \& Primack 1999; Efstathiou 2000; Thacker \& Couchman 2001; 
Springel \& Hernquist 2003) have been phenomenological in nature, as they 
contain one or more free parameters that need to be adjusted to 
match observations, e.g., the observed rates of star formation. 
What is lacking is a detailed high-resolution 3D model of 
turbulent flows within a small region of the ISM, to help 
identify basic properties of interstellar turbulence. 

Our aim in this paper is to provide a framework in which to handle SN 
feedback in cosmological simulations. We divide this challenging task into 
two parts. First, we run a suite of numerical simulations of 
stratified ISM driven by discrete SN explosions.  
The SN rate and the vertical gas column density are systematically varied over wide ranges 
to examine how the velocity dispersion changes as a function of 
the gas density, length scale, and the assumed SN rate. 
In particular, we study the distribution and variation of turbulent pressure 
in these models. 
These local, high-resolution simulations have high enough spatial resolution 
to follow turbulent motions of the interstellar gas in detail. 

The second step is to use the local ISM models to find a prescription for 
the physics unresolved in global, cosmological simulations so that 
we have a better idea of how to handle SN feedback. 
We propose a subgrid model based on heuristic analytic arguments and 
simulation results. The idea is 
as follows: If we average relevant physical quantities over scales greater 
than the largest energy containing scale, we may describe the large-scale 
dynamics of the ISM that is insensitive to the details of the small-scale 
physics. This is motivated by similar studies in the past (Yepes et al.\ 
1997; Springel 2000) as well as the new finding in Paper I that $>$90\% of 
the total kinetic energy of the disk is contained on scales below 
200 pc, comparable to the size of resolution elements in current 
cosmological simulations (e.g., Governato et al.\ 2004, 2007; 
Cen et al.\ 2005; Naab et al.\ 2007; Gibson et al.\ 2008; Joung et al.\ 2008).  
This suggests the 
interesting possibility that the SN-driven turbulence may be handled 
separately as subgrid physics in those simulations. 
Based on the physical characteristics 
of our local ISM models with discrete SN explosions, the subgrid model is 
physically well-motivated. 
We apply the results of our ISM models by formulating an 
accurate feedback prescription 
after deriving an effective equation of state for turbulent pressure of the medium. 
Such a prescription may be used 
to examine the effect of SN feedback in large cosmological volumes.

Dib et al.\ (2006) investigated the relationship between 
the velocity dispersion of the gas and the SN rate in 3D simulations 
of unstratified, periodic boxes with a fixed mean density.  
Their aim was to explore the 
observed constancy of the velocity dispersion in the outer parts of galactic disks 
(Dickey et al.\ 1990; Kamphuis 1993) and the transition to the starburst regime.  
They simulated a (1 kpc)$^3$ volume of the 
unstratified ISM with 128$^3$ cells for various SN rates, while keeping the gas 
density of the box constant. 
They found nearly constant velocity dispersions for 
the \ion{H}{1} gas at relatively low SN rates (0.01 to 0.5 times the Galactic value; 
note thtat they took a Galactic SN rate per volume of 
$2.58 \times 10^2$ Myr$^{-1}$ kpc$^{-3}$), 
as well as a transition to the starburst regime at about the star formation rate 
per unit area of 0.5--1$\times$10$^{-2}$ M$_{\odot}$ kpc$^{-2}$ yr$^{-1}$ 
(assuming a Salpeter initial mass function; Salpeter 1955), 
in good agreement with observations. 
The present study extends their work to stratified 
atmospheres at higher spatial resolution where the gas density is 
systematically increased with increasing SN rate.

The basic features of our ISM models are described in \S~\ref{models}. 
In \S~\ref{genchar}, we describe the thermal and structural properties of the models.  
We demonstrate in \S~\ref{turbpres} that the turbulent pressure of a medium with 
a given SN rate is independent of the local density, and study how the 
velocity dispersion and \ion{H}{1} linewidth vary 
as a function of the assumed SN rate. Using these results, we construct a new 
subgrid model and calibrate it against the ISM simulations 
in \S~\ref{subgrid}. 
Finally, we discuss several outstanding open issues 
(\S~\ref{discuss}) and summarize our main results (\S~\ref{summary}).

\section{Local ISM Simulations}
\label{models}
 
We use the model described in Paper I and its extensions described here. 
Our simulations are performed using Flash v2.4, an 
Eulerian astrophysical hydrodynamics code with adaptive mesh refinement 
(AMR) capability, developed by the Flash Center at the University of 
Chicago (Fryxell et al.\ 2000). It solves the Euler equations using the piecewise-parabolic 
method (Colella \& Woodward 1984) to handle compressible flows with shocks. 
For parallelization, the Message-Passing Interface library is used; 
the AMR is handled by the PARAMESH library.

Although, in the present work, we consider 
SNe as the only source of interstellar turbulence, other processes 
such as gravitational instability (e.g., Li et al.\ 2005a), 
magneto-Jeans instability (Kim \& Ostriker 2002), 
magnetorotational instability (e.g., Piontek \& Ostriker 2005), 
expanding \ion{H}{2} regions (Matzner 2002) 
and protostellar outflows (Haverkorn et al.\ 2004; 
Nakamura \& Li 2007; Banerjee et al.\ 2007; Carroll et al.\ 2008) should 
all contribute to the observed turbulence on various length scales.
However, Mac Low \& Klessen (2004) argue that these other processes 
may play relatively minor roles.  
We assume that they are 
insignificant on the spatial scales resolved in our models.

Our computation box contains a volume of (0.5 kpc)$^2 \times$(10 kpc), 
elongated in the vertical direction. 
If the grid were fully resolved at our maximum resolution of 1.95 pc, 
it would contain 256$^2 \times$5120 zones, but fully refined cells are mostly 
concentrated near the galactic midplane (the central $\pm 200$ pc) of our models. 
The gravitational potential of Kuijken \& Gilmore (1989) is 
employed.  For simplicity, the gravitational potential is fixed 
for all models, 
   except in the model with the highest SN rate (512x), where the increased 
   gas surface density dominates the gravitational potential, as it 
   exceeds the assumed stellar mass surface density of $4.46 \times 10^7$ 
   M$_{\odot}$ kpc$^{-2}$ (Binney \& Tremaine 1987).  Assuming the Schmidt-Kennicutt 
   law and taking $\Sigma_{\rm tot} = 75$ M$_{\odot}$ pc$^{-2}$ and $\Sigma_{\rm gas} 
   = 5.3$ M$_{\odot}$ pc$^{-2}$ for the solar neighborhood (Binney \& Tremaine 1987), 
   we increase the gravitational acceleration due to the disk component 
   in the Kuijken \& Gilmore potential by a factor of 11 for the 512x model.  
   This is our standard 512x model, and the results are based on this model, 
   unless otherwise noted.  For comparison, we also ran a model with the same 
   SN rate but with the gravitational potential used in the lower SN rate models; 
   this latter model will be denoted 512xL.  
Self-gravity of gas is not included.  
Outflow boundary conditions are used in the upper and 
lower surfaces parallel to the Galactic plane, while periodic 
boundary conditions are employed elsewhere.

For each SN explosion, we dump thermal energy $E_{\rm SN} 
= 10^{51}$ ergs in a small sphere whose radius varies as a 
function of the local density. The radii are chosen such that 
radiative losses inside the spheres are negligible in the first 
few timesteps after explosions (Paper I); 
they usually range between $\sim$7 
and $\sim$50 pc. 
We use them to trace the thermal history of metal particles and to 
estimate their escape fraction from the galaxy. We assume that a 
fixed fraction, 3/5, of the Type II SNe 
are closely correlated in space and in time as a way of simulating superbubbles. 
The remaining explosions have random positions scattered through 
the galaxy mass, with scale heights of 325 pc for Type I and 90 pc 
for Type II SNe, to represent field SNe.  
The 
   minimum number of SNe $n_{\rm sn,min} = 7$, while the
    average and maximum numbers of SNe per superbubble $\langle n_{\rm sn} \rangle$
    and $n_{\rm sn,max}$
are determined by a power-law distribution 
$dN_{\rm B} \propto n_*^2 \, dn_*$ 
   yielding the values given in Table~\ref{tbl_models}
(Kennicutt et al.\ 1988; see also Paper I 
for details).  
We fix the Type I SN rate at the Galactic value 
   $\Sigma_{\rm *,I} =6.58$ Myr$^{-1}$ kpc$^{-2}$ in all models, 
as they take $>$1 Gyr to explode. 

\begin{deluxetable}{c|crrcc}
\tablecaption{Parameters for the ISM simulations\label{tbl_models}}
\tablewidth{0pt}
\tablehead{
\colhead{model}  & \colhead{$\dot{\Sigma}_{\rm *,II}$ $^a$} & \colhead{$\Sigma_{\rm gas}$ $^b$} & \colhead{$\Gamma$ $^c$} & \colhead{$n_{\rm sn,max}$} & \colhead{$\langle n_{\rm sn} \rangle$}
}
\startdata
1x           & 11.0, 16.4   & 1.87(6) & 8.50({-26})     &    40 & 14.8 \\
8x           & 98.0, 147.   & 8.23(6) & 1.55({-25})     &   320 & 27.4 \\
64x          & 794., 1190.  & 3.63(7) & 2.78({-25})     &  2560 & 41.4 \\
512x         & 6360., 9540. & 1.61(8) & 5.11({-25})     & 20480 & 55.9 \\
\enddata
\tablecomments{The SN rates per unit area are in units of Myr$^{-1}$
  kpc$^{-2}$. Exponents are given parenthetically.}
\tablenotetext{a}{$\;$ Rates for (field, 
association) Type II SNe}
\tablenotetext{b}{$\;$ The gas column density at $t=0$ is given in M$_{\odot}$ kpc$^{-2}$}
\tablenotetext{c}{$\;$ The diffuse heating rate is expressed in erg s$^{-1}$}

\end{deluxetable}

We run a suite of numerical models with wide ranges of SN rates 
and gas column densities, as listed in Table \ref{tbl_models}. 
We adopt SN rates per unit area, 
$\dot{\Sigma}_*$, of 1, 8, 64, and 512 times the Galactic rate 
(see Table \ref{tbl_models} for the exact values adopted), and 
vary the gas column densities systematically 
following the Schmidt-Kennicutt law (Kennicutt 1998). 
Wong \& Blitz (2002) reported observational 
evidence for seven CO-bright spiral galaxies that the azimuthally 
averaged star formation rate per unit area scales with the gas column 
density similarly to the disk-averaged star formation law (see Li et al.\ 
2006 for similar results from numerical simulations). 
More recently, Kennicutt et al.\ (2007) reported evidence that the 
same star formation law is satisfied on scales of 0.5--2 kpc. 
The initial gas column densities $\Sigma_{\rm gas}$ are determined by 
inverting the relation  
\begin{equation}
\label{sigmastar}
\dot{\Sigma}_* = A \, \left( \frac{\Sigma_{\rm gas}}{1 \; {\rm M}_{\odot} \; {\rm pc}^{-2}} \right)^{1.4} \, . 
\end{equation}
where the coeffcient $A = 2.5\times10^{-4} \; {\rm M}_{\odot} \; {\rm yr}^{-1} \; {\rm kpc}^{-2}$.  
As the Schmidt-Kennicutt law holds in various physical conditions 
found in galaxies, we believe that our choices of ($\Sigma_{\rm gas}$, $\dot{\Sigma}_*$) 
are representative of typical regions in actual galaxies.

Radiative cooling appropriate for an optically thin 
plasma with $Z$$=$$Z_{\odot}$ (Dalgarno \& McCray 1972; Sutherland 
\& Dopita 1993; see Fig. 1 of Paper I) and a diffuse heating term 
that accounts for the photoelectric heating (Wolfire et al.\ 1995, 2003) 
of the neutral gas are also included.  
The diffuse heating rate $\Gamma$ in each model 
may not be directly proportional to $\dot{\Sigma}_*$ because  
the far-ultraviolet (FUV) radiation responsible for the photoelectric heating 
($h \nu = 6$--$13$ eV) 
may be significantly shielded in regions of high densities.  
In reality, the FUV radiation field is spatially nonuniform and time-varying 
(Parravano et al.\ 2003).  
However, as radiative transfer is not included in our calculations, 
we resort to a simple scaling relation.  
To compute the mean intensity of the FUV radiation appropriate for each ISM model, 
we imagine a medium that emits {\it and} absorbs radiation uniformly.  
In the optically thick limit 
(appropriate for our box size of 500 pc), the intensity of 
the FUV radiation, $I_{\rm FUV} = j_{\nu}/\alpha_{\nu} \,$, 
where $j_{\nu}$ is the emissivity and $\alpha_{\nu} = n \sigma$ is the absorption 
coefficient with the number density of absorbers, $n$, and their cross-section, $\sigma$.  
Assuming 
$j_{\nu} \propto \dot{\Sigma}_* \propto \Sigma_{\rm gas}^{\, 1.4} \,$ 
and $n \propto \Sigma_{\rm gas}$ (since the gas scale height 
$H_{\rm gas}$ is only weakly dependent on the SN rate, as shown later), we
obtain $I_{\rm FUV} \propto \Sigma_{\rm gas}^{\, 0.4}\propto \dot{\Sigma}_*^{\, 2/7}$.  
With the additional assumption that 
$\Gamma$ is proportional to the intensity of the FUV radiation, we obtain 
$\Gamma \propto \dot{\Sigma}_*^{2/7}$.  
As in Paper I, the diffuse heating rate in each model declines exponentially 
away from the galactic midplane with a scale height of 300 pc.

We note that 
another set of models was run with a different scaling relation for 
the diffuse heating rate: 
$\Gamma \propto \dot{\Sigma}_* \propto \Sigma_{\rm gas}^{1.4} \,$.  
These models, however, yielded results inconsistent with observations.  
For example, almost all the gas 
in the 512x model in this set resided in the warm phase 
($T \sim 10^4$ K, the maximum gas temperature 
where diffuse heating is applied in our models), and the turbulent pressure 
was smaller than the thermal pressure by a factor of a few, 
in disagreement with observations.  
Our result is consistent with the simulations in Kim (2004), who 
found that the velocity dispersion of the gas in a SN-driven medium is 
lower for higher external pressure. 
Although we do not present these models in this paper, they 
illustrate the importance of including a proper form of diffuse heating rate, 
as it affects the balance among various interstellar gas phases. 

\section{Thermal and Structural Properties}
\label{genchar}

\begin{figure*}
\epsscale{1.0}
\hspace{-10mm} \plotone{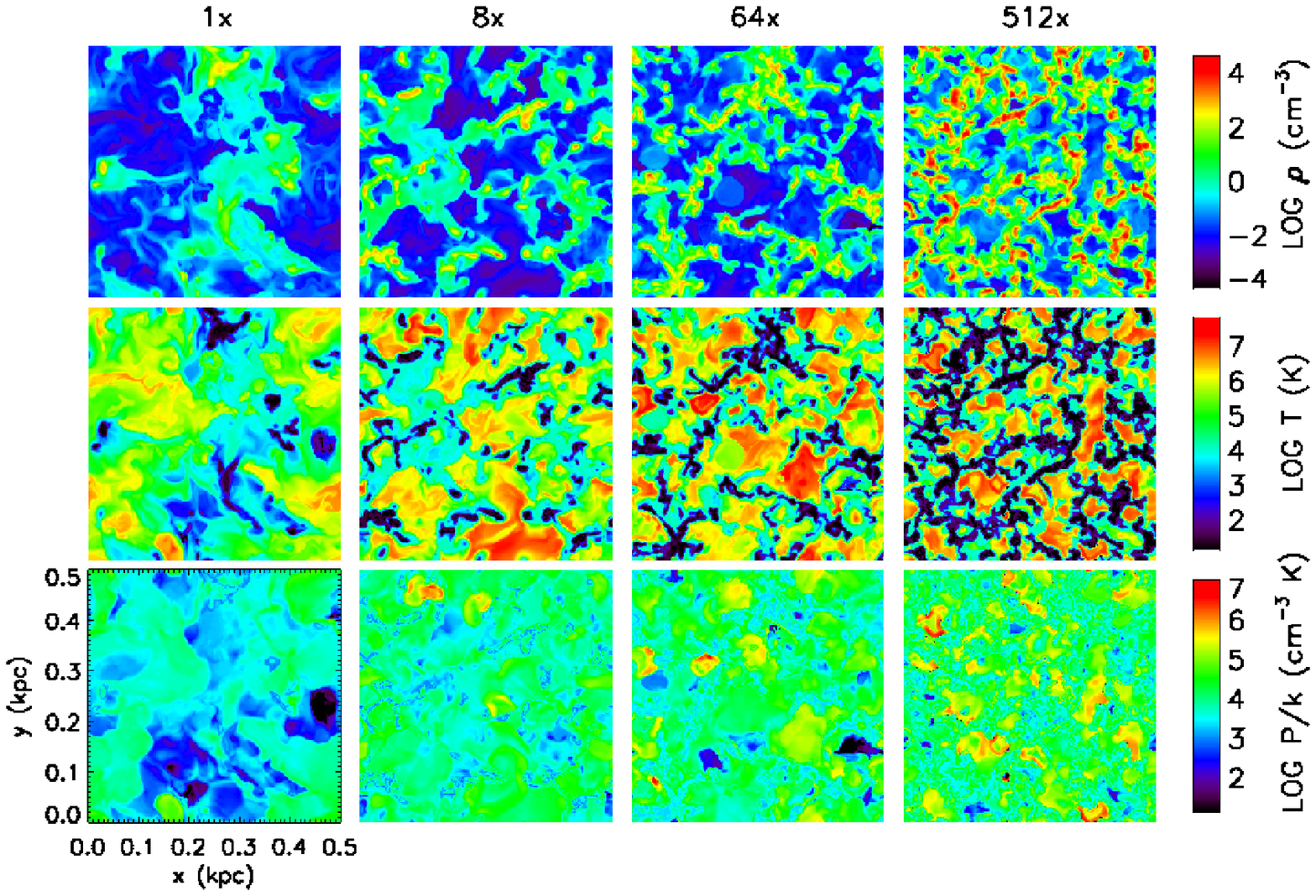}
\caption{{\it (From top to bottom)} gas density, temperature, and thermal pressure in the 
midplane of our four ISM models. Each column corresponds to a single model as indicated 
at the top of each column: {\it (from left to right)}, 1, 8, 64, and 512 
times the Galactic SN rate.  The colorbars shown on the right apply to 
all models.  
\label{mid_all}}
\end{figure*}

\begin{figure}
\epsscale{1.0}
\begin{center}
\plotone{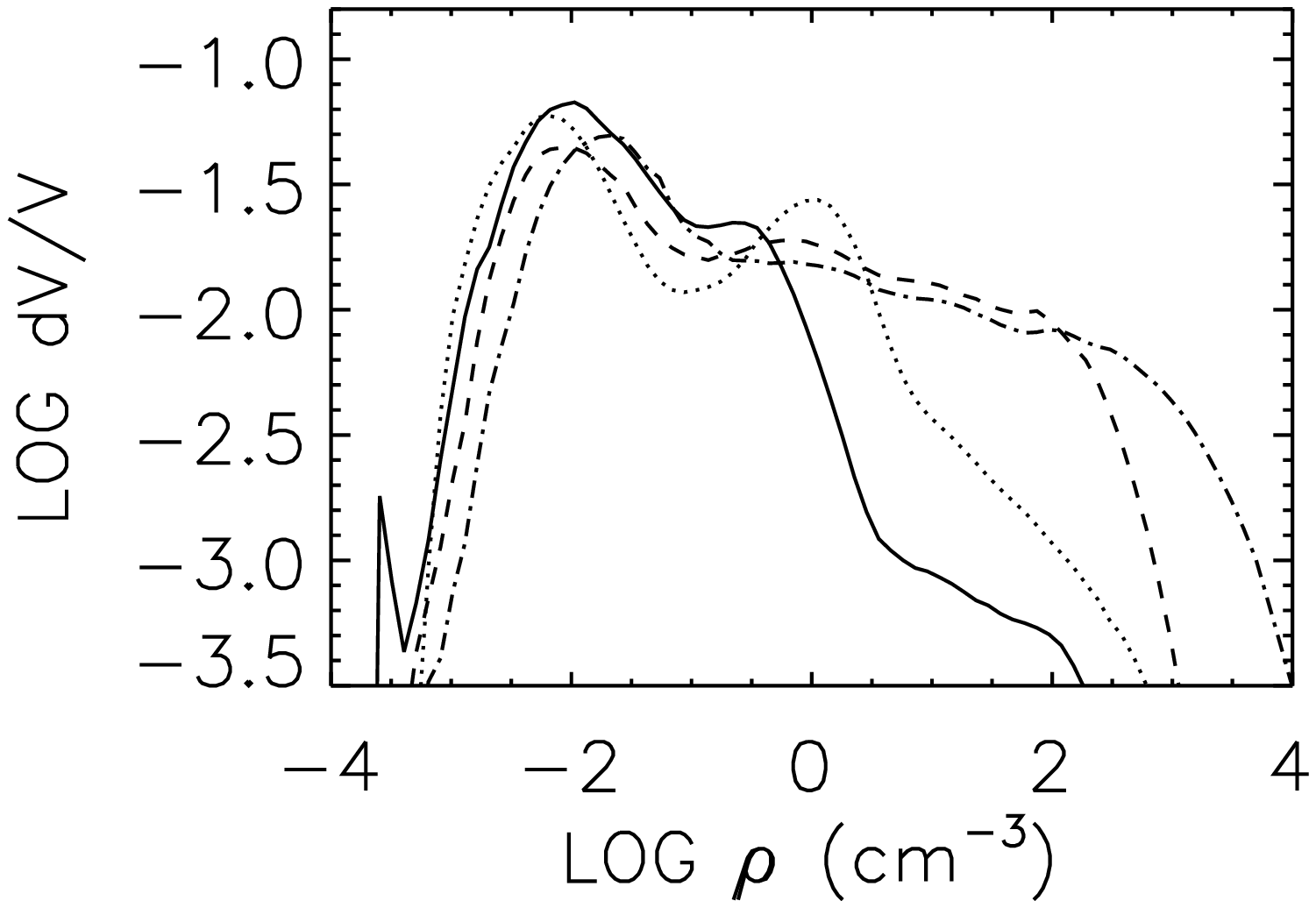}

\vspace{-10mm}
\plotone{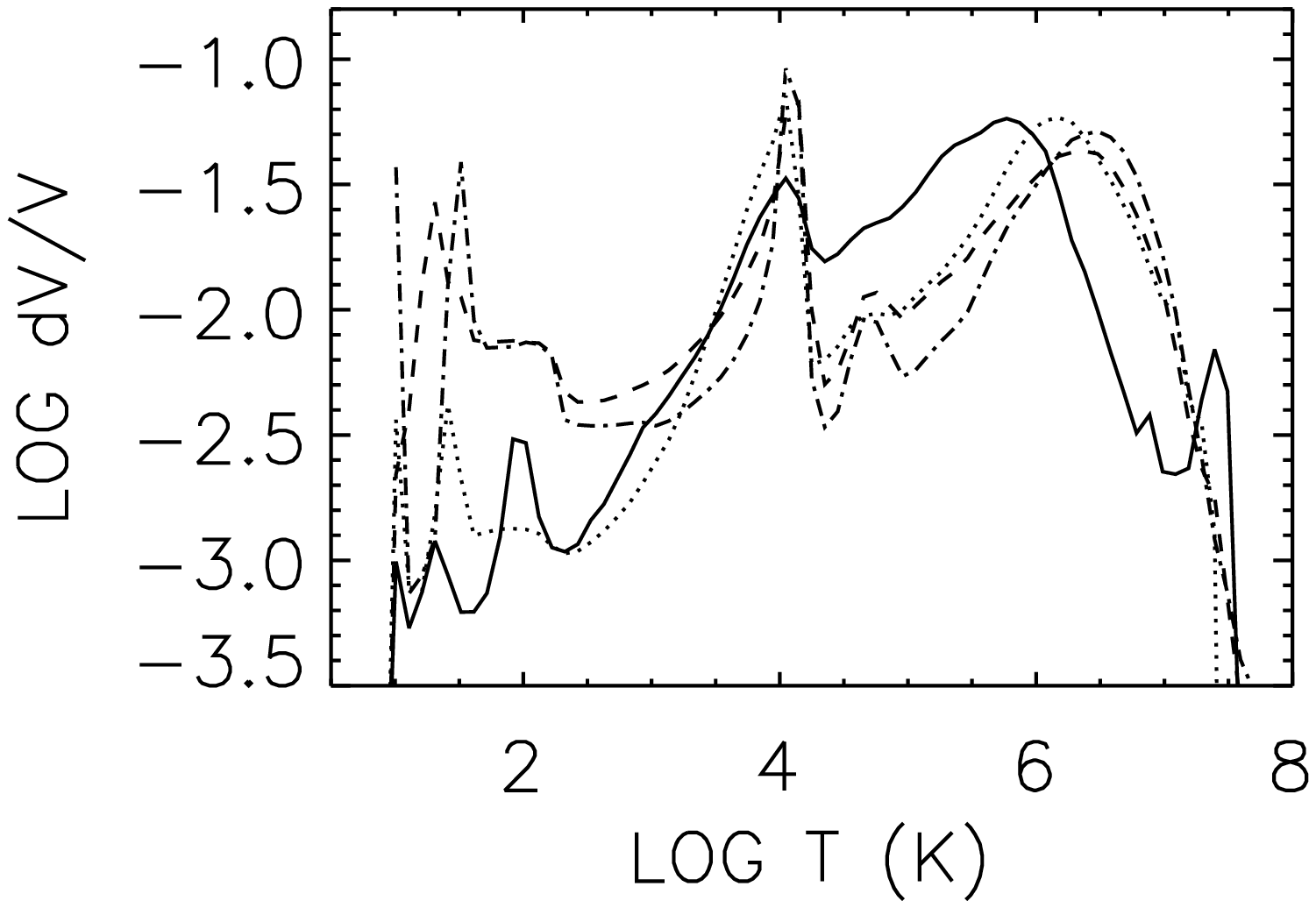}

\vspace{-10mm}
\plotone{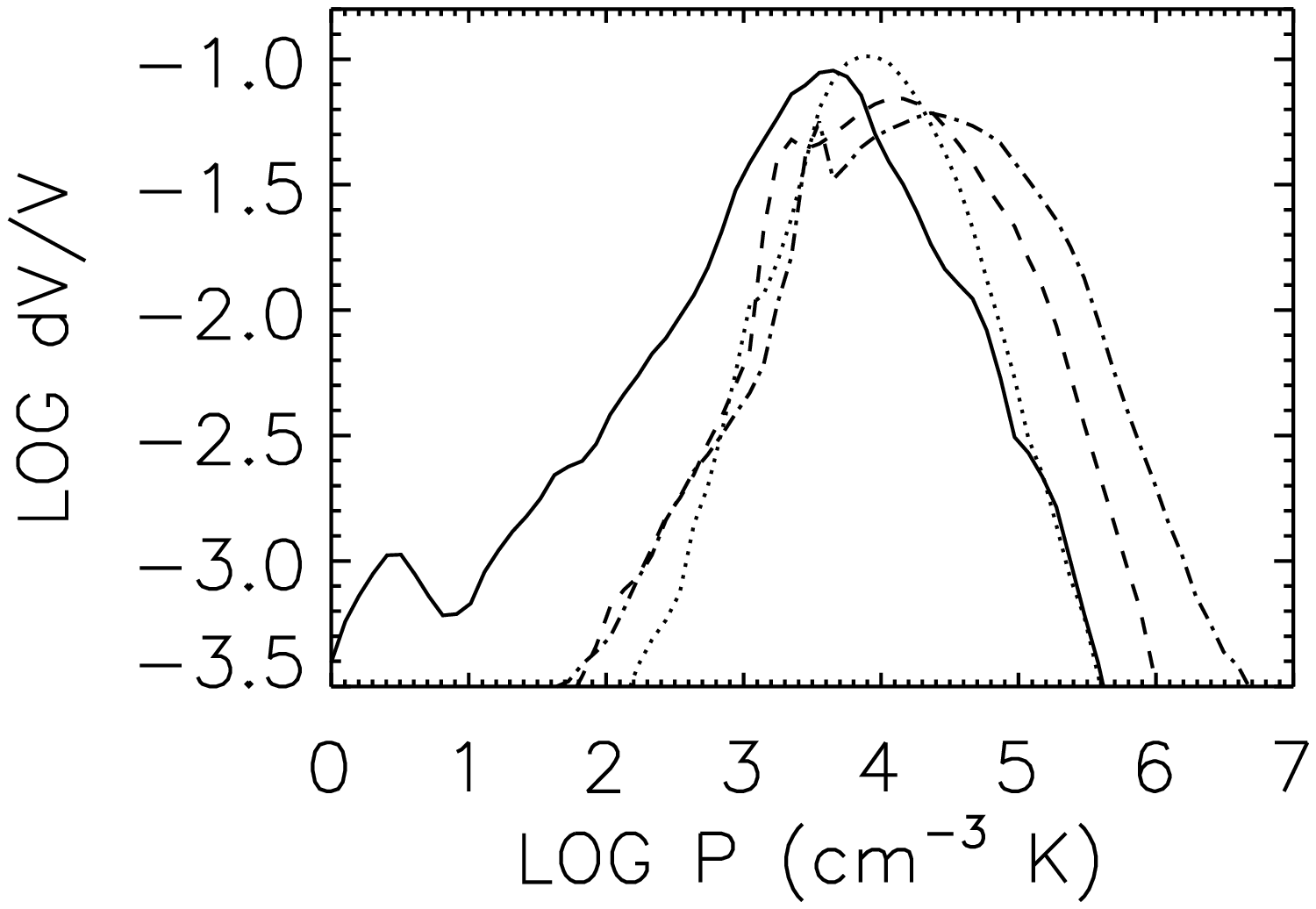}
\end{center}
\caption{(From top to bottom) Probability distribution functions of 
(a) gas density, (b) temperature, and (c) thermal pressure near the 
midplane ($|z| \le 125$ pc) of all models showing fractional volumes 
contained in logarithmic bins.  The 1x, 8x, 64x, and 512x models are 
shown in solid, dotted, dashed, and dot-dashed curves, respectively.
\label{vfrac}}
\end{figure}

Figure \ref{mid_all} displays (from top to bottom) gas density, 
temperature, and thermal pressure in the midplane of the four ISM models. 
Figure \ref{vfrac} shows 
probablity distribution functions for those variables.  
As in the Galactic SN rate model discussed in Paper I, 
the three classical phases of the ISM are present in every model, 
with cold, dense filaments surrounded by warm gases that in turn 
are embedded in hot media.  
However, their volume fractions vary as a function of the SN rate:  
The hot and cold phases increase in volume with increasing SN rate, 
while the volume occupied by the warm phase generally decreases (Fig. \ref{vfrac}b).  
Interestingly, the gas fraction in the thermally unstable range 
of temperatures (200 $\lesssim T \lesssim$ 7000 K) increases 
steadily with the SN rate (Fig. \ref{vfrac}b), reflecting more dynamic 
ISM with higher SN rates.  
More subtle differences are also found.  For example, 
the mean density of the hot gas becomes progressively higher as the SN rate 
increases (Fig. \ref{vfrac}a), likely caused by an increased level of turbulent diffusion 
off dense shells and filaments where SN explosions occur more frequently 
(de Avillez \& Mac Low 2002).  
Partly due to the increase in density, the overall thermal pressure 
$P_{\rm ther}$ in the midplane 
increases with the SN rate (Fig. \ref{vfrac}c).  
Within each ISM model, however, we find that 
thermal pressure varies within a range of roughly two orders of magnitude 
(Mac Low et al.\ 2005).  
The peak of the pressure distribution in the Galactic SN rate model lies at 
$\sim$$3 \times 10^3$ cm$^{-3}$ K, close to the mean value measured with C I 
fine-structure lines (Jenkins \& Tripp 2007). 
Recent SN remnants are associated with thermal pressures higher than the mean, 
whereas old SN remnants are tied to those that are lower.  
The dispersion in $P_{\rm ther}$ rises for high SN rate models, 
where its maximum value reaches $\sim$$10^7$ cm$^{-3}$ K, close to the 
value measured near the center of M82 (Strickland \& Heckman 2007). 
Our results are broadly in agreement with the analytic model by Monaco (2004) 
in which the ISM was modeled as a two-phase medium in pressure equilibrium.

\begin{figure}
\epsscale{0.61}
\hspace{-17mm} \plotone{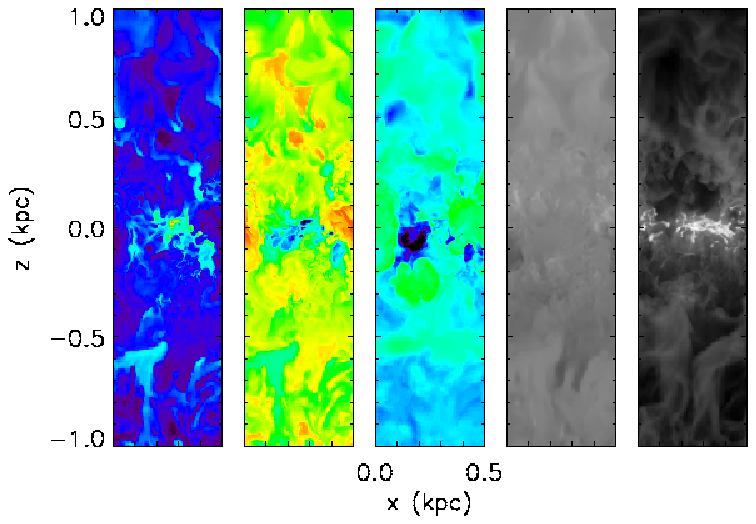} 
\hspace{-4mm} \plotone{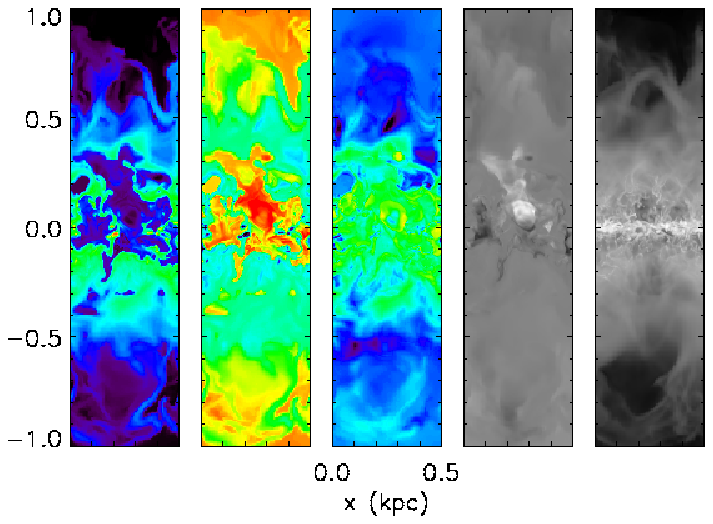}

\vspace{-5mm}
\hspace{-17mm} \plotone{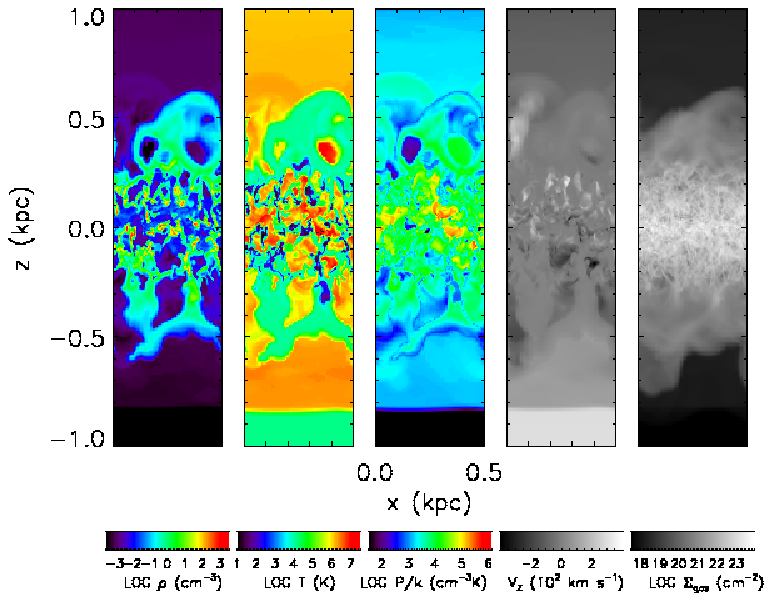} 
\hspace{-4mm} \plotone{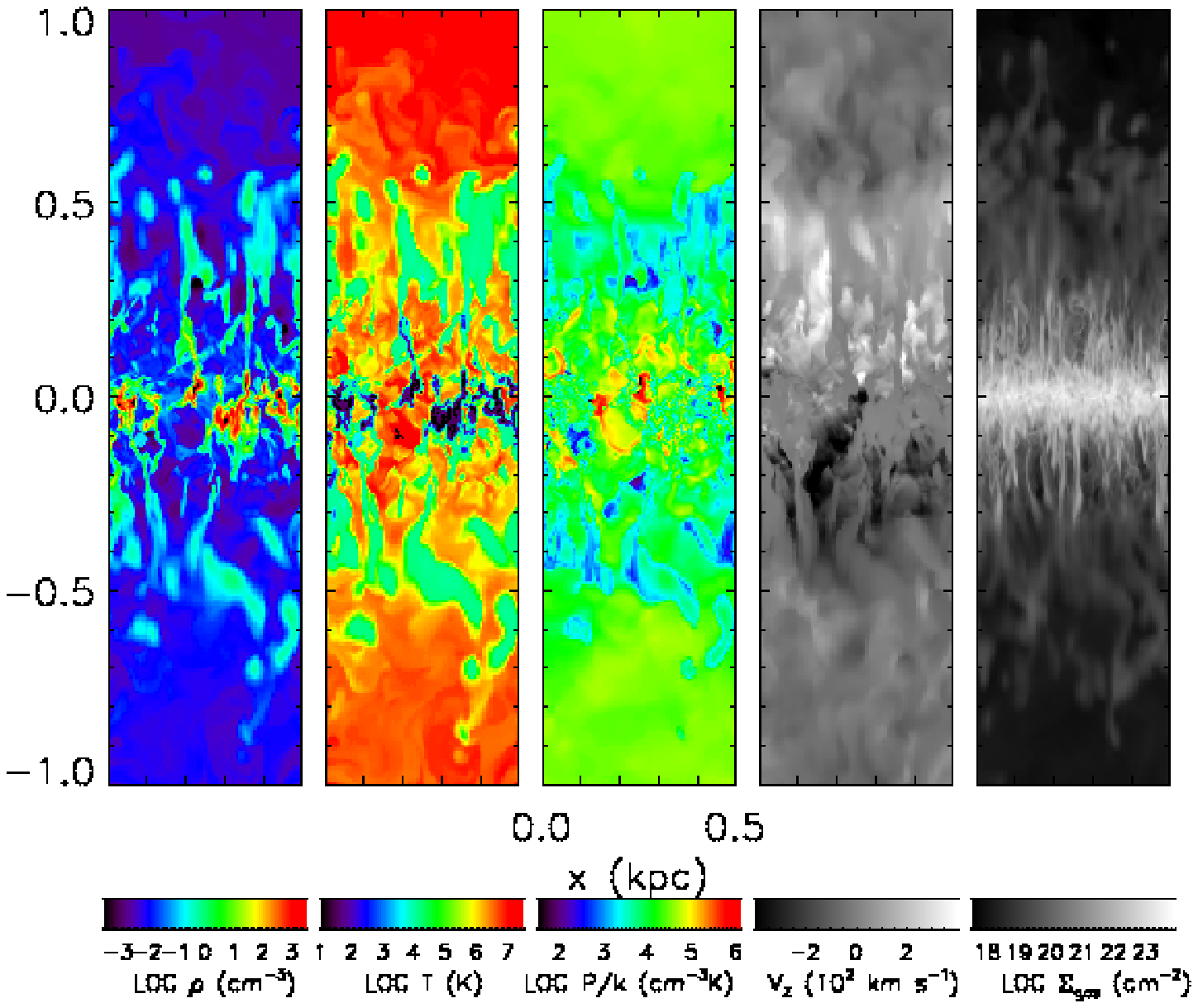}
\caption{Vertical distributions of the four ISM models in this paper, 
(a) 1x, (b) 8x, (c) 64x, and (d) 512x, from a snapshot in time. 
For each model, ({\it from left to right}) slices of 
gas density, temperature, thermal pressure, and z-component of the velocity, 
and the projected gas column density are shown for the middle 2 kpc of the 
models ($|z| \le 1$ kpc). 
\label{vert_all}}
\end{figure}

The vertical distributions of the ISM models are shown in Figure \ref{vert_all}.
In all models, correlated SN explosions create superbubbles that vent hot 
($T \gtrsim 10^6$ K) metal-enriched gas out of the galactic disk into the halo with 
high velocities ($|v_{\rm z}| \gtrsim 300$ km s$^{-1}$).  
As the SN rate increases, the total gas mass also increases; a gradually larger 
mass fraction resides in the cold phase rather than in the warm phase.
With stronger gravity, we would expect the dense gas to be more 
confined to the galactic plane, decreasing the disk thickness, while 
superbubbles would blow out more easily into the halo.  
On the other hand, the change in the ISM near the galactic midplane should be 
modest.  
    This is what we find by comparing the two models, 512x and 512xL.  
    The 512x model has higher gas densities in the midplane as well as higher 
    turbulent velocities than those in the 512xL model.  In particular, it is 
    worth noting that 
    the 512x model shows superbubble blow-outs, contrary to expectations 
    from some previous work (see \S~\ref{discuss} for more details).

\section{Turbulent Pressure Distribution}
\label{turbpres}

We examine the distribution of turbulent pressures in our models after 
the systems reach statistical steady states at $t \approx 80$ Myr. 
Because kinetic energy is distributed over a broad range of wavelengths 
(see Fig. 8b of Paper I), turbulent pressure is inherently a scale-dependent 
quantity. To quantify the random (i.e., turbulent) component of the velocity 
field apart from the bulk motion of the medium, 
the root-mean-square (rms) velocity dispersion $\sigma$ is measured in the 
local frame of reference, i.e., the center of mass frame of a selected 
volume. We choose 
the entire plane with 
$|z| \leq 125$ pc, and tile it with small cubical boxes, starting 
from those with only 2 zones (3.91 pc) on a side. The velocity 
dispersion within each box is computed from the random component of 
the velocity field $\bw$ after subtracting the center-of-mass velocity 
of the box $\bov_0$ from the original velocity field $\bov$. The box 
size is successively increased twofold 
until it reaches 128 zones (250 pc) on a side. This procedure enables us to 
study the velocity dispersion as a function of scale. For each box of a given 
size, the turbulent rms velocity dispersion $\sigma_{\rm turb,b}$ is 
calculated by
\begin{equation}
\sigma_{\rm turb,b}^2 = \frac{1}{M_{\rm b}} \int_{\rm box} \rho \, |\bw|^2 \, dV \; ,
\end{equation}
where $M_{\rm b}$ denotes the total mass within the box and 
$\bw \equiv \bov - \bov_0$. 
Similarly, the thermal velocity dispersion $\sigma_{\rm ther,b}$ is computed by 
replacing $|\bw|$ with the local sound speed $c_{\rm s}$. We prefer to use quantities 
directly related to pressure, so we set 
\begin{equation}
\label{sigther}
\sigma_{\rm ther} = (\sigma_{\rm ther,b}^2/\gamma)^{1/2} \, , 
\end{equation}
\begin{equation}
\label{sigturb}
\sigma_{\rm turb} = (\sigma_{\rm turb,b}^2/3)^{1/2}\, , 
\end{equation}
and 
\begin{equation}
\label{sigtot}
\sigma_{\rm tot} = (\sigma_{\rm ther,b}^2/\gamma+\sigma_{\rm turb,b}^2/3)^{1/2} 
\end{equation}
for 1D thermal, turbulent, and total velocity dispersions, respectively 
(see \S~\ref{discuss}). 
We hereafter drop the subscript ``b'' with the 
understanding that all velocity dispersions are scale-dependent quantities, 
computed for boxes of a particular size.  Similarly, we define the total pressure 
$P_{\rm tot}$ as the sum of thermal ($P_{\rm ther}=nkT$) and turbulent ($P_{\rm turb}=\rho 
\, \sigma_{\rm turb}^2$) pressures. Below we report our findings on the distributions 
of turbulent pressure found from this analysis. 

\subsection{Pressure Equilibrium at Fixed SN Rate}
\label{pequil}

\begin{figure}
\epsscale{1.25}
\plotone{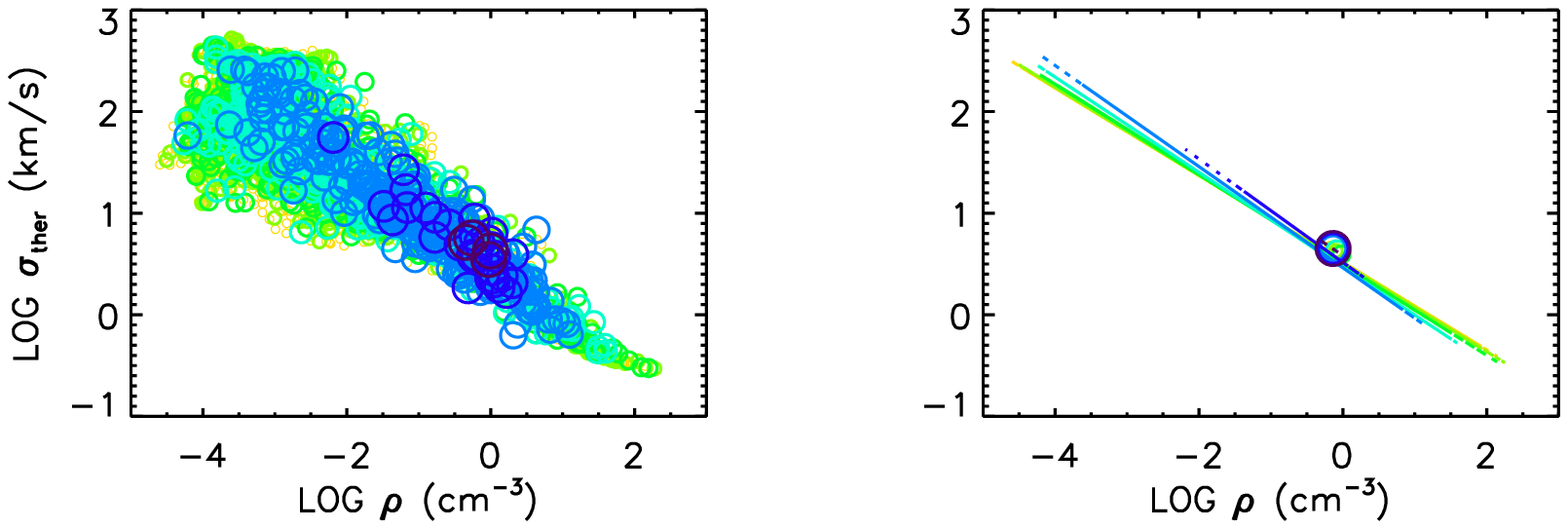}
\plotone{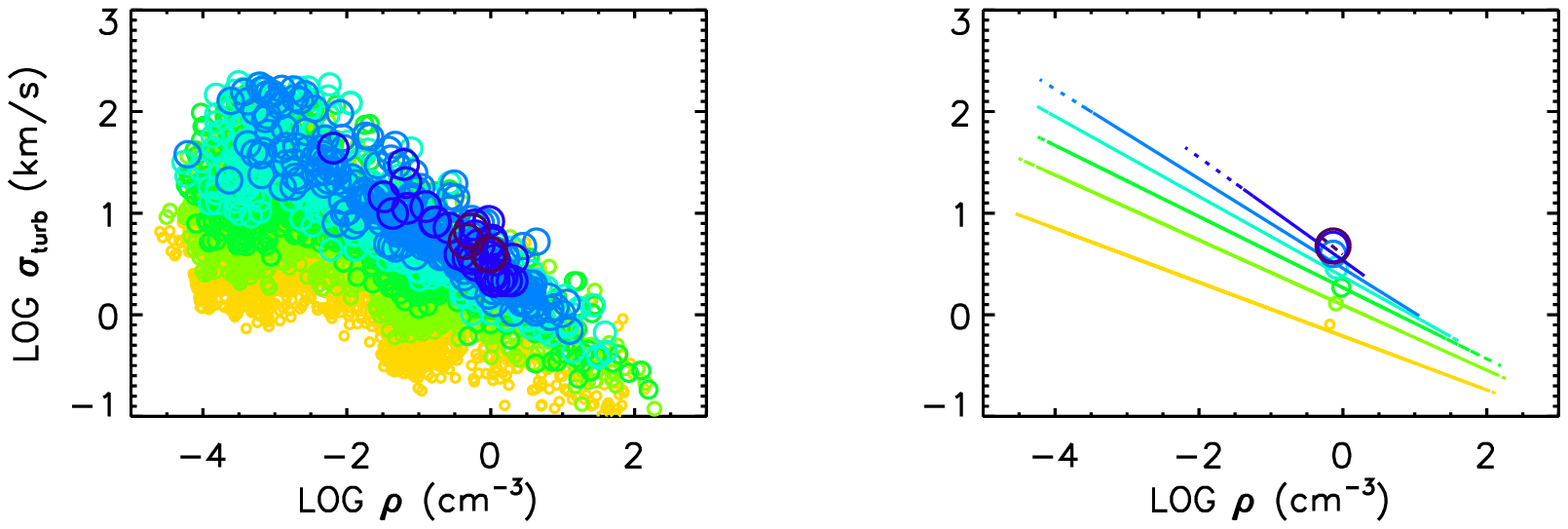}
\plotone{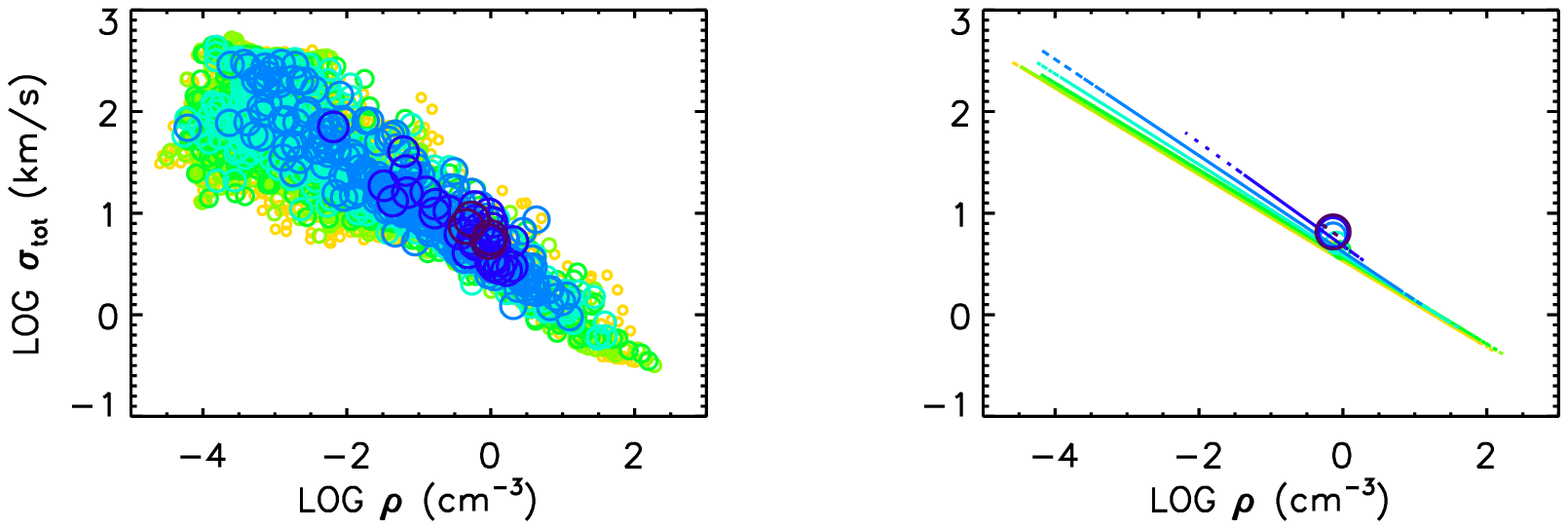}
\plotone{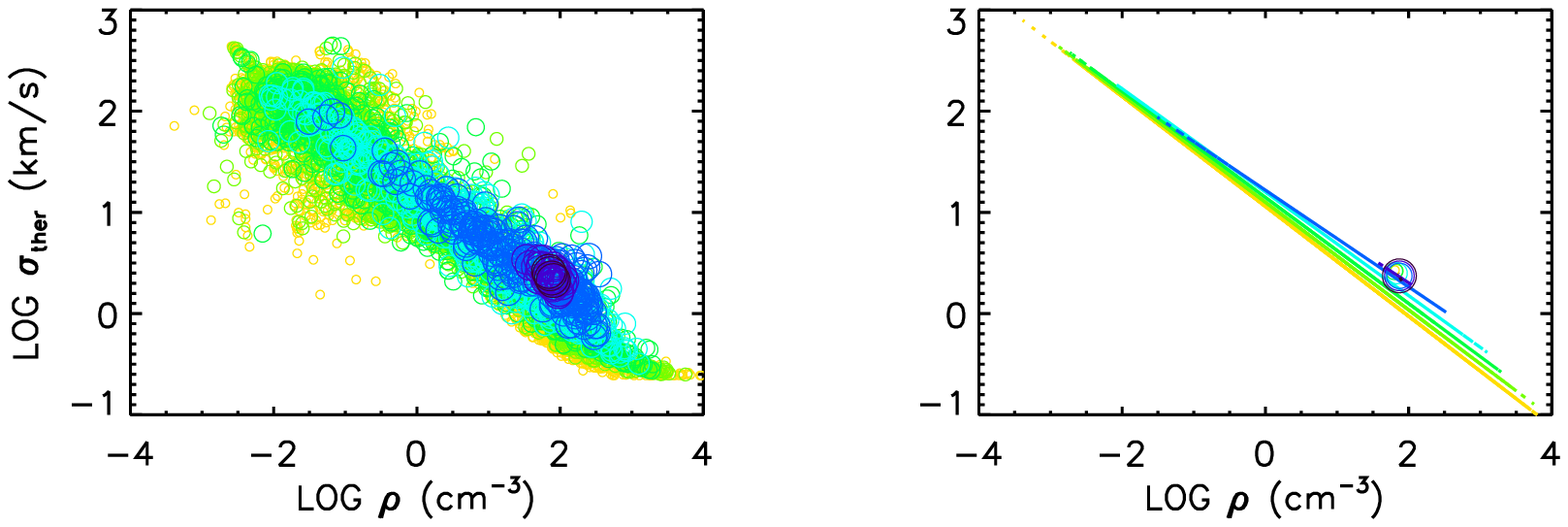}
\plotone{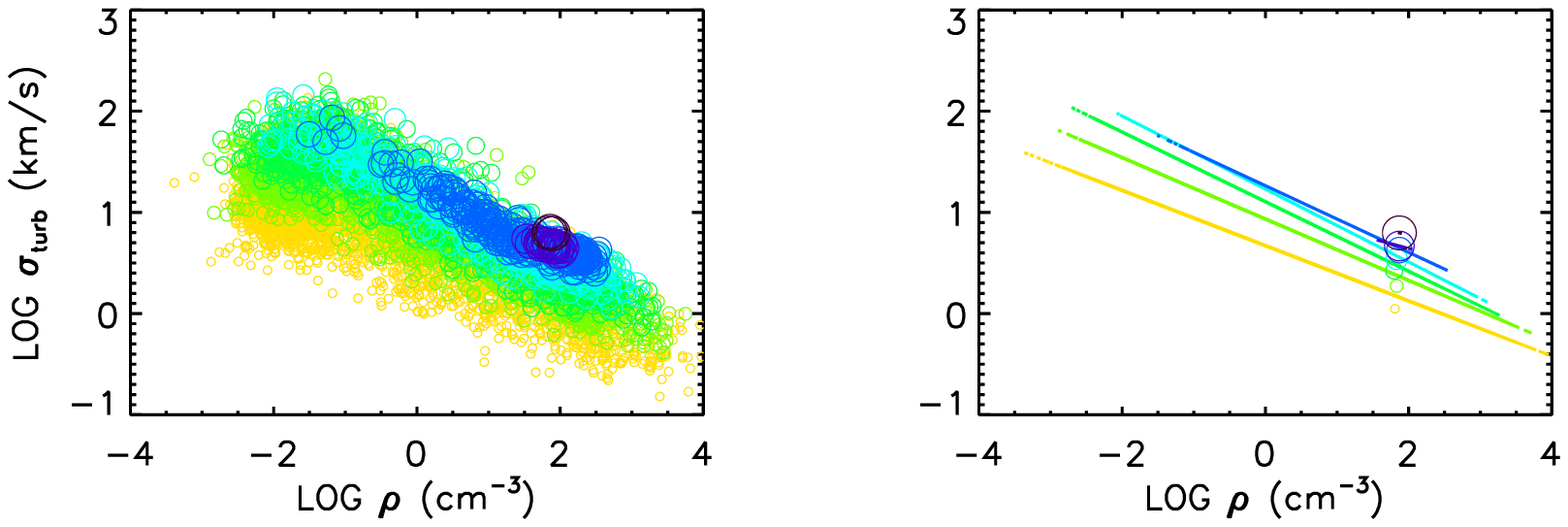}
\plotone{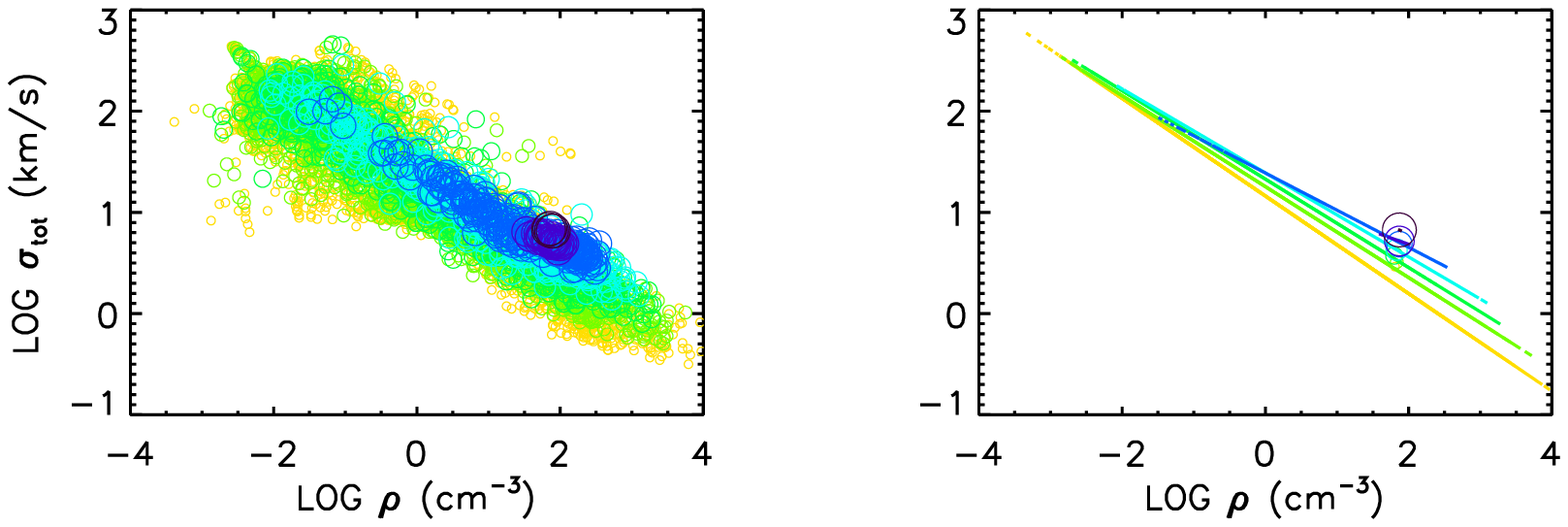}
\caption{Mass-weighted (a) thermal, (b) turbulent, and (c) total 
velocity dispersions plotted against average densities of the boxes. The data 
are taken near the midplane of our Galactic SN rate model for various box 
sizes over which the velocity dispersions were computed. Small yellow corresponds to 
the smallest cubic boxes with 3.91 pc on a side, and purple to the largest boxes 
with 125 pc on a side. The plots on the 
right-hand-side display linear regressions of the corresponding plots in the left 
column, row by row. On a log-log plot, the distribution of points can be 
characterized by a slope of $-1/2$, implying a nearly constant total pressure. 
In (b) and (c), the velocity dispersion increases with scale because, as
the box size increases, larger eddies gradually contribute to the turbulent 
velocity dispersion. (d)--(f) are similar plots for the 512x model.
\label{sigs}}
\end{figure}

In Figure \ref{sigs}, thermal, turbulent, and
total velocity dispersions (Eq. \ref{sigther}--\ref{sigtot}) 
are plotted against average densities of individual boxes 
as a function of the box size. The plots on the 
right-hand-side display linear regressions of the corresponding plots in the left 
column, row by row. Small yellow circles represent 
the smallest boxes (3.91 pc on a side), while big purple ones represent 
the largest boxes (250 pc on a side). The results are presented only 
for the 
1x and 512x
Galactic SN rate models at $t \approx 80$ Myr, 
but we verified that 
the conclusions in this section hold for all four models 
throughout their steady-state evolution.

The most striking feature of these plots is that, for low and 
intermediate densities 
($n \lesssim 10$ cm$^{-3}$) the mean velocity dispersion, 
$\bar{\sigma}
\propto \rho^{\, -1/2}$, showing {\em isobaric} 
behavior.  This is true at every scale for  
both thermal and total velocity dispersion, and also at large scales for 
the turbulent velocity dispersion. The idea of {\it thermal} pressure 
balance between phases is not at all new (Field et al.\ 1969). Indeed, McKee \& 
Ostriker (1977)'s three-phase ISM model was also based on the premise of 
``rough pressure balance'' between phases, motivated in part by 
Chandrasekhar \& Fermi (1953) and Spitzer (1956). 
What is new and remarkable is that the {\it turbulent} pressure is 
likewise in approximate pressure equilibrium (Fig. \ref{sigs}b) 
on large scales where $P_{\rm turb} \gtrsim P_{\rm ther}$. 
In particular, at any given scale, $\sigma_{\rm tot} 
\propto \rho^{\, -1/2}$, indicating that the total pressure $P_{\rm tot} = \rho \, 
\sigma_{\rm tot}^2$ is nearly constant in the midplane.  
This implies that different gas tracers, which often probe 
narrow ranges of gas densities, may find discrepant kinematics 
(in particular, velocity dispersions) but that there may be an 
underlying correlation between them. 
The pressures are in approximate equilibrium, although 
the gas density and temperature individually vary by $\sim$7 orders of magnitude. 

Furthermore, the thermal velocity dispersion is scale-independent, 
as indicated by the lines in the top-left panel of 
Figure \ref{sigs}, which lie almost on top of one another. In contrast, 
the turbulent velocity dispersion increases with scale. As the 
box size increases, larger eddies gradually contribute to the turbulent 
velocity dispersion, and $\sigma_{\rm turb}$ increases 
just as the power spectrum of kinetic energy 
predicts (\S~\ref{kespec}). As a result, the thermal pressure dominates 
the total pressure on small scales, 
while the turbulent pressure dominates 
on large scales. At any particular scale, the scatter is about an order of magnitude in 
$\sigma$, leading to $\sim$2 orders of magnitude scatter in the pressure (see Fig. 
7c of Paper I).


\subsection{Turbulent Velocity Dispersion vs. SN Rate}
\label{sigma}

\subsubsection{Mass-weighted velocity dispersion}
\label{mwveldisp}

\begin{figure}
\epsscale{1.2}
\plotone{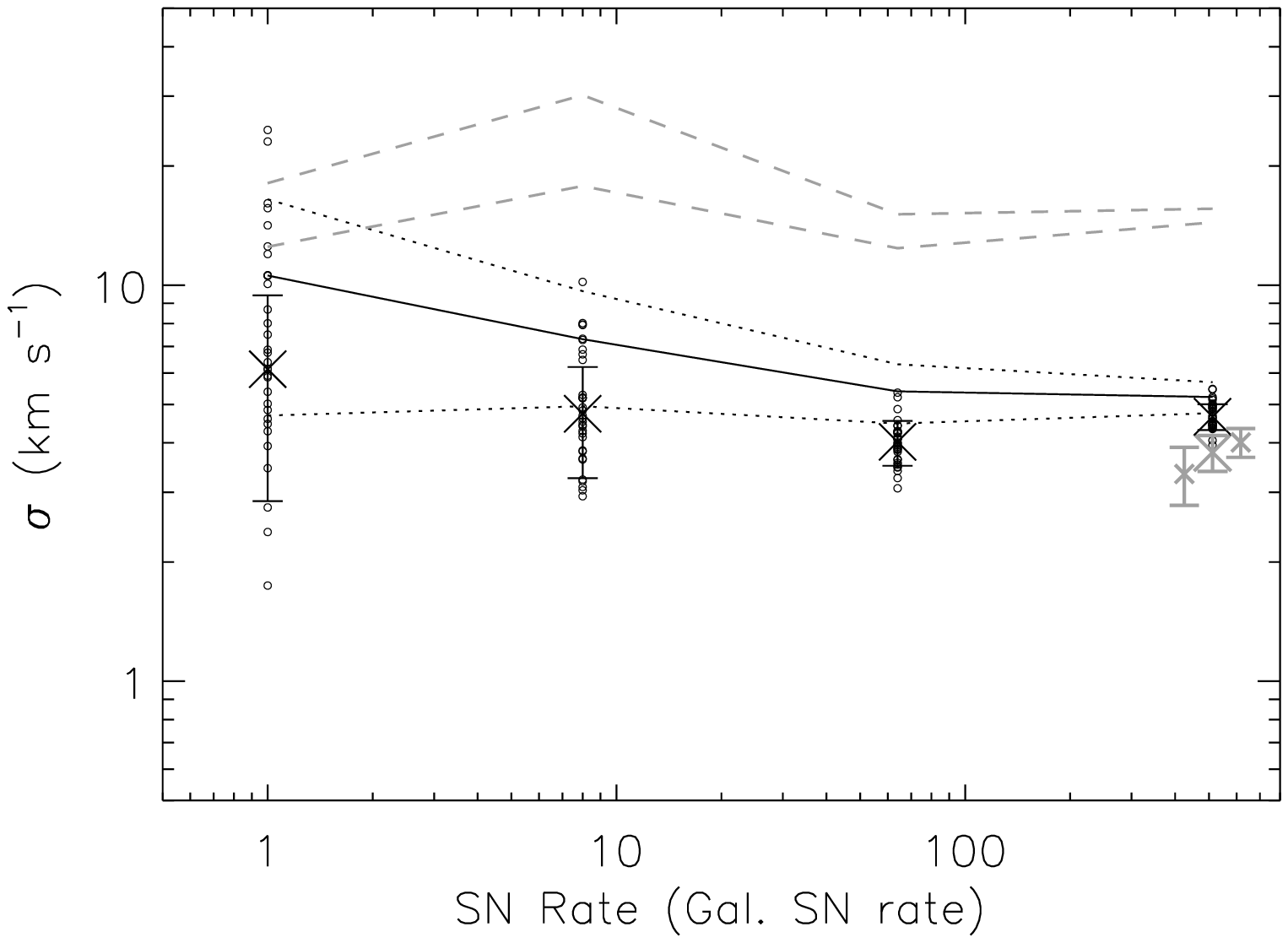}
\caption{Turbulent 1D 
velocity dispersions in boxes with 125 pc sides near the midplane 
(open circles), plotted against 
the SN rates in our four models. 
The changes in the mean values (represented by the X's and error bars) 
are minimal across the range of 512 in SN rates.  
The points in thick grey lines indicate 
turbulent velocity dispersions measured in the 512xL models 
with spatial resolutions lower (3.9 pc/cell; {\it left}) and 
higher (0.98 pc/cell; {\it right}) than the fiducial model (1.95 pc/cell; {\it middle}), 
offset along the x-axis for clear presentation. 
The variation of the mean {\em total} (thermal plus turbulent) 
1D velocity dispersions are shown 
in solid line, bounded by one sigma dispersions (dotted lines).  
The grey dashed lines represent the linewidths of \ion{H}{1} emission, 
computed with the entire line profiles ({\it upper curve}) and 
with only the line centers ({\it lower curve}), as described in the text.
\label{vel_disp}}
\end{figure}

How does turbulent pressure scale with the SN rate? To answer this 
question, we compare $\sigma_{\rm turb}$ of the second largest boxes with 125 pc 
on a side in all models 
(purple circles in Fig. \ref{sigs}b) and obtain relevant physical quantities 
averaged over 125 pc scales.  
(Hence, unless otherwise noted, $\sigma_{\rm turb}$ and $P_{\rm turb}$ 
henceforth refer to the values computed in boxes with 125 pc sides.)  
Using these models, we can find how turbulent
pressure depends on the SN rate, or after conversion via an appropriate IMF, the 
star formation rate.

Figure \ref{vel_disp} displays the 1D mass-weighted 
turbulent velocity dispersion against the 
SN rate for the four SN rates and gas column densities 
employed in our models. 
The mean velocity dispersions
$\sigma_{\rm turb}$ (in the mass-weighted rms sense) 
and their 
standard deviations are given in Table~\ref{tbl_results}
The changes in the mean values 
are minimal across the range of 512 in SN rates:  
the best-fit line has the form 
\begin{equation}
\label{sigmafit}
\sigma_{\rm turb} = (5.6 \pm 0.8) \, \dot{\Sigma}_{*,1}^{\; \, -0.045 \, \pm \, 0.033} \; \, {\rm km \; \rm s}^{-1} \, .  
\end{equation}
The variation of the mean {\em total} (i.e., thermal plus turbulent) 
1D velocity dispersions is shown 
in solid line, bounded by one sigma dispersions (dotted lines).  
For the mean values for this quantity
    given in Table~\ref{tbl_results}
the 3D velocity dispersions are 8--18 km s$^{-1}$, 
comparable to the sound speed of the warm gas. 
The scatter in the velocity dispersion decreases with increasing SN rate.

\begin{deluxetable}{c|ccccc}
\tablecaption{Results from the ISM simulations\label{tbl_results}}
\tablewidth{0pt}
\tablehead{
\colhead{model} & \colhead{$\sigma_{\rm turb}$} & \colhead{$\sigma_{\rm tot}$} & \colhead{$\sigma_{\rm v,H~I}$ $^a$} & \colhead{$L_{\rm d,eff}$ $^b$} & \colhead{$f_{\rm w}$}
}
\startdata
1x    & $6.1\pm3.3$ & $10.6\pm5.9$ & 12.5 & 137 & 0.460 \\
8x    & $4.7\pm1.5$ &  $7.3\pm2.4$ & 17.8 &  87 & 0.274 \\
64x   & $4.0\pm0.5$ &  $5.4\pm0.9$ & 12.4 &  57 & 0.431 \\
512x  & $4.7\pm0.4$ &  $5.2\pm0.5$ & 14.4 &  55 & 0.609 \\
\enddata
\tablecomments{See text for definitions of the variables.}
\tablenotetext{a}{$\;$ The first three columns are in units of km s$^{-1}$.}
\tablenotetext{b}{$\;$ In pc}
\end{deluxetable}

For a resolution study, we ran the 512xL model 
with spatial resolutions half (3.91 pc/cell; left) and 
double (0.98 pc/cell; right) that of the fiducial model (1.95 pc/cell; middle).  
The turbulent velocity dispersions measured in these models 
are indicated by the error bars in thick grey lines without open circles, 
offset along the x-axis for clear presentation. 
The velocity dispersions are 
$3.3\pm0.6$, $3.8 \pm 0.4$, and $4.0\pm0.3$ km s$^{-1}$ for the low, intermediate 
and high resolution models, 
respectively.  Therefore, although the 512xL models have not fully converged, 
the velocity dispersion is likely converging to a finite value 
at infinite resolution, 
since factor of two changes in resolution led to only 
$\sim$24\% and $\sim$5\% increases in 
the velocity dispersion, with rapidly declining fractional changes.  
Since the 512xL model is associated with the highest mean density and 
hence most affected by resolution-dependent effects, 
we expect better convergence in the other three models.  

\subsubsection{\ion{H}{1} linewidths}

High-resolution \ion{H}{1} emission maps of 
nearby, almost face-on galaxies are now available 
(e.g., Petric \& Rupen 2007; Tamburro et al.\ 2008).  
These observations show how the \ion{H}{1} line widths and shapes 
change as a function of the galactocentric distance, gas density, 
distance from spiral structure, 
and star formation rate.  Hence, they provide important constraints on the 
dominant driving mechanism of the neutral gas.  

The brightness temperature, $T_b$, 
is computed by (e.g., Spitzer 1978) 
\begin{equation}
T_b \, (w) = (5.49 \times 10^{-19} \mbox{ K}) \left( \frac{N_{{\rm H~I}}}{\Delta w} \right) \left( \frac{1-e^{-\tau_w}}{\tau_w} \right) \, ,
\end{equation}
where $w$ is the velocity offset from the 21 cm line center, $N_{{\rm H~I}}$ is the 
column density of H~{\sc i} in cm$^{-2}$, $\Delta w$ is the velocity width 
in km s$^{-1}$ that 
we take to be 2.5 km s$^{-1}$, which was the velocity resolution in 
Petric \& Rupen (2007), and the velocity-dependent 
optical depth is given by 
\begin{equation}
\tau_w = 5.49 \times 10^{-19} \left( \frac{N_{{\rm H~I}} \; e^{-(w/b)^2}}{\pi^{1/2} \, b \, T} \right) \, ,
\end{equation}
where the Doppler linewidth $b = (2kT/m)^{1/2}$ in units of km s$^{-1}$.   
Since we do not follow the multiple species of hydrogen explicitly in our 
simulations, our estimate of $N_{{\rm H~I}}$ is based on simple 
temperature cutoffs.  
Specifically, we assume that 100\% of hydrogen between 50 and 7000 K 
exists as \ion{H}{1}, while the neutral fraction increases linearly from 30 to 50 K.  
Below 30 K, hydrogen is assumed to be fully molecular.  
Note that the \ion{H}{1} linewidths are generally larger than the 
turbulent velocity dispersions.  This happens because the bulk 
of the highest density gas, associated with the lowest velocity 
dispersion, is molecular and so excluded in the calculation of 
the former quantity.
The uncertainty in \ion{H}{2} fraction and in H$_2$ fraction affect 
the line profiles in the broad wings and in the centers, respectively. 

\begin{figure}
\epsscale{0.57}
\plotone{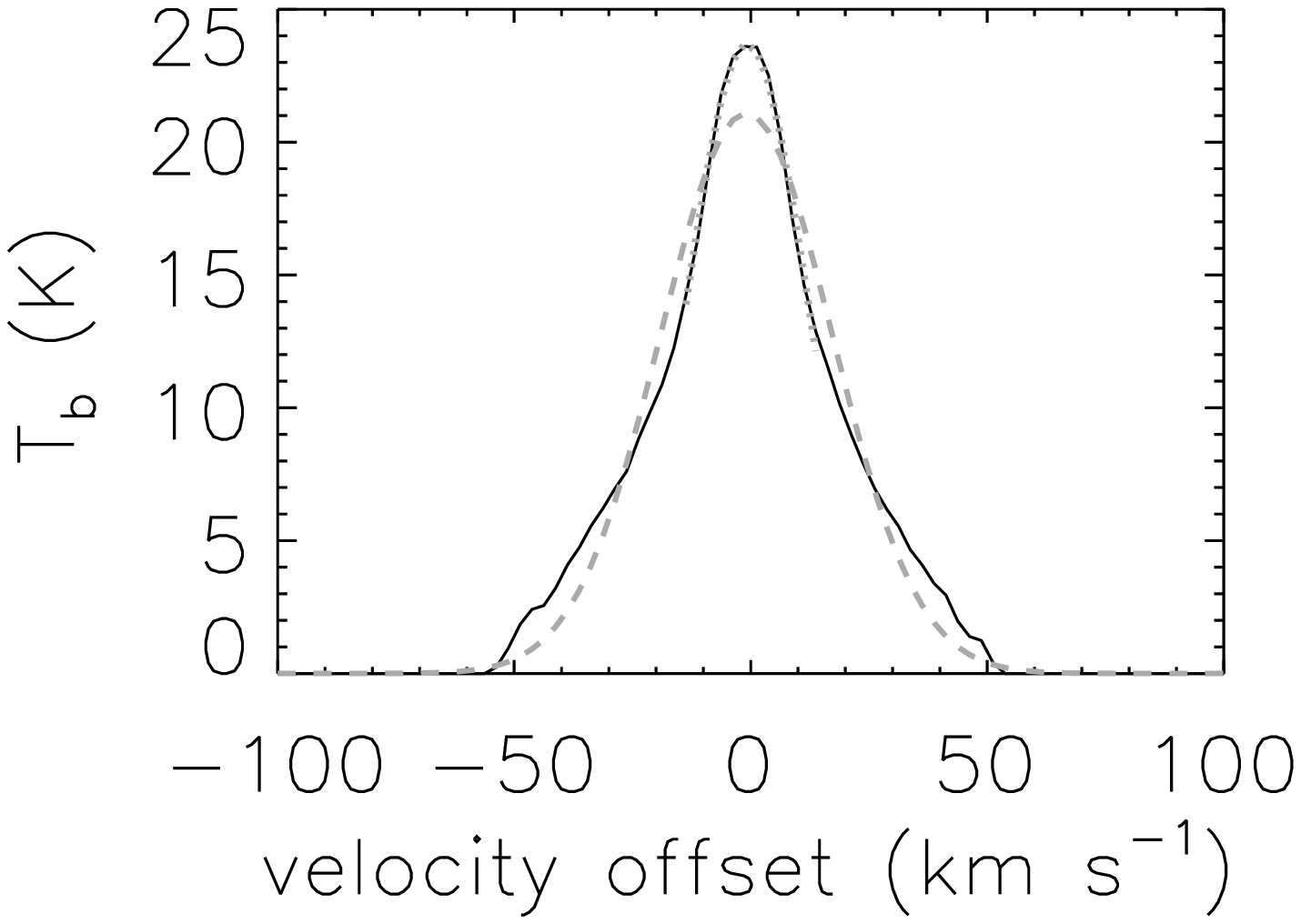}
\plotone{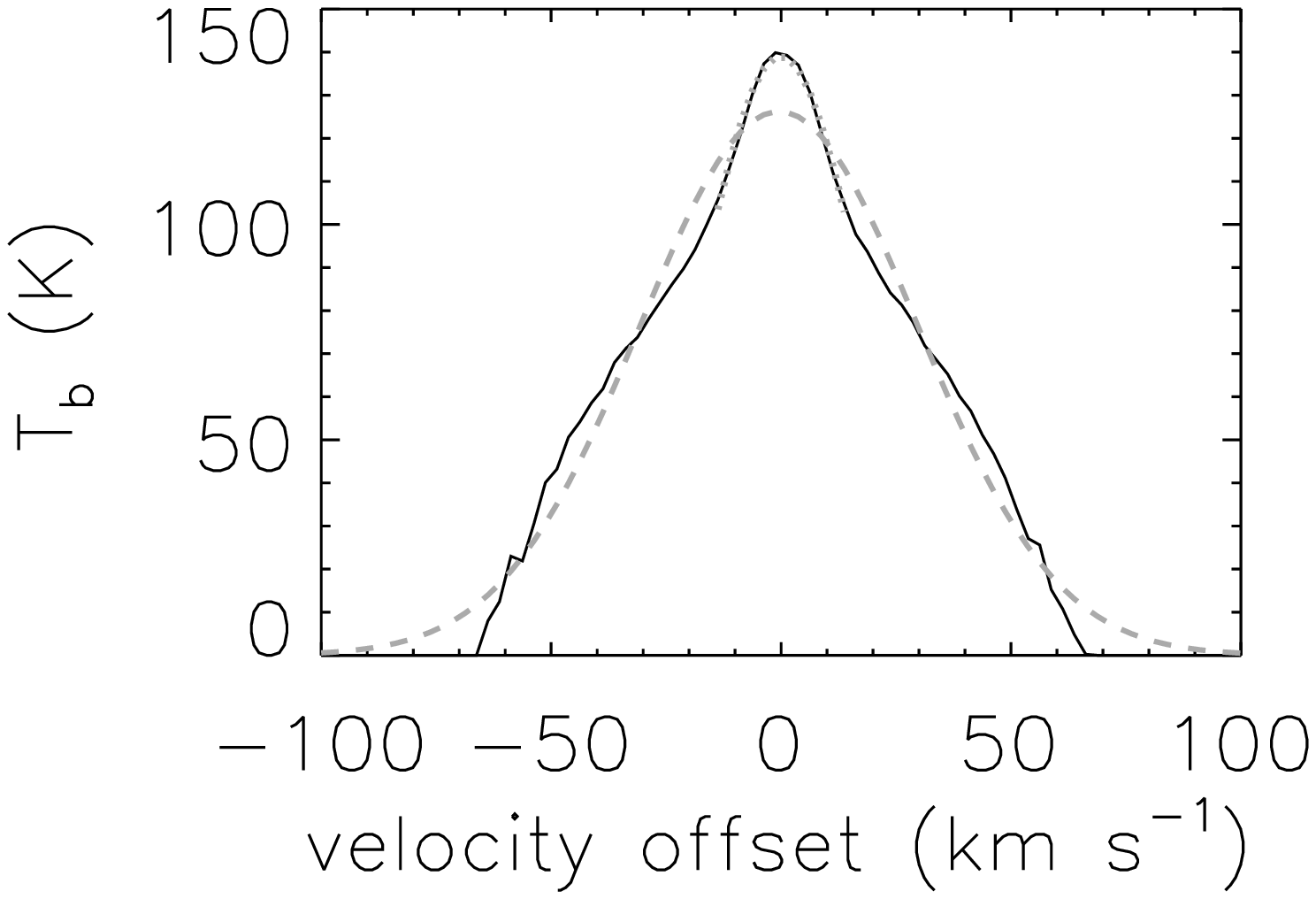}
\plotone{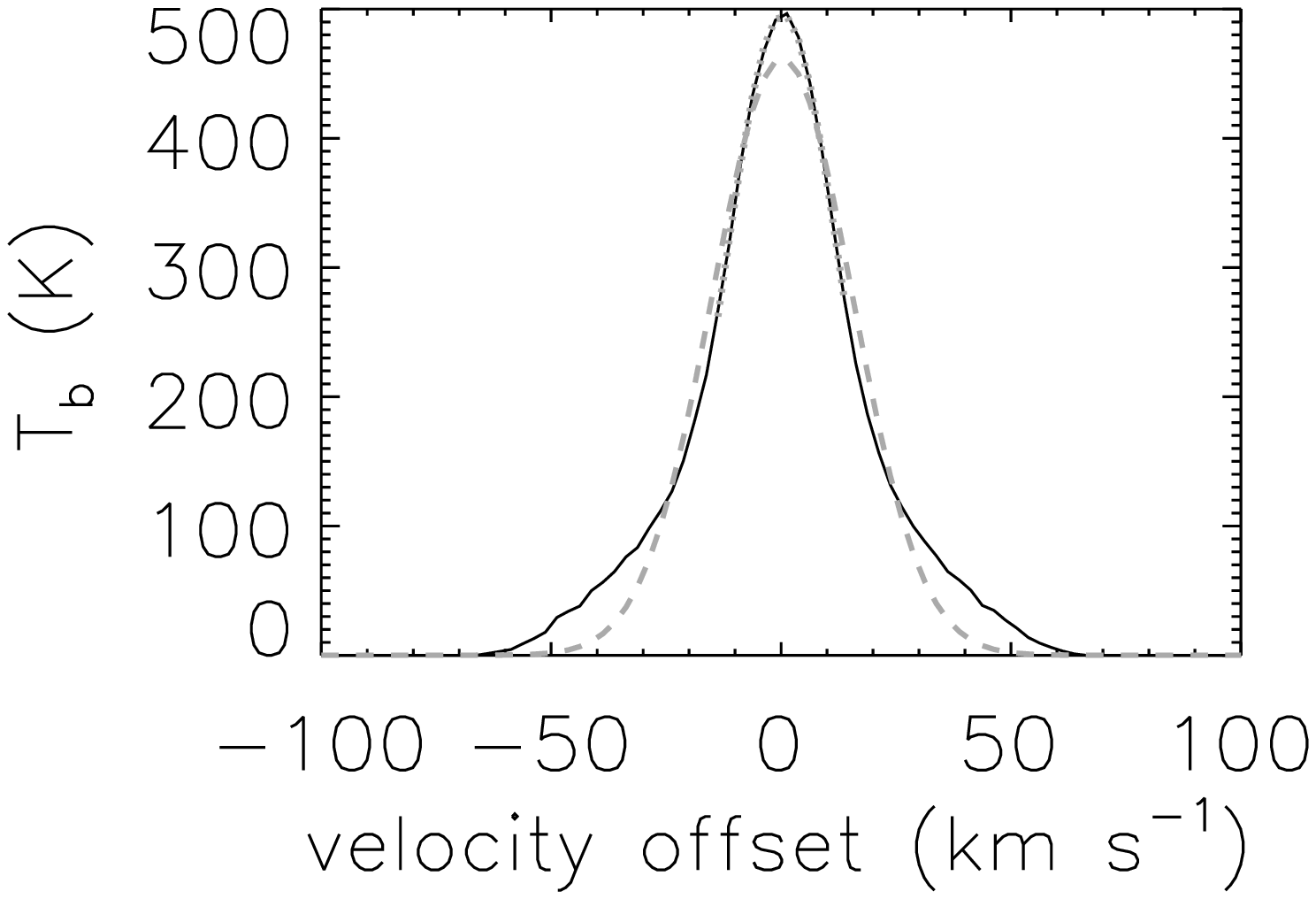}
\plotone{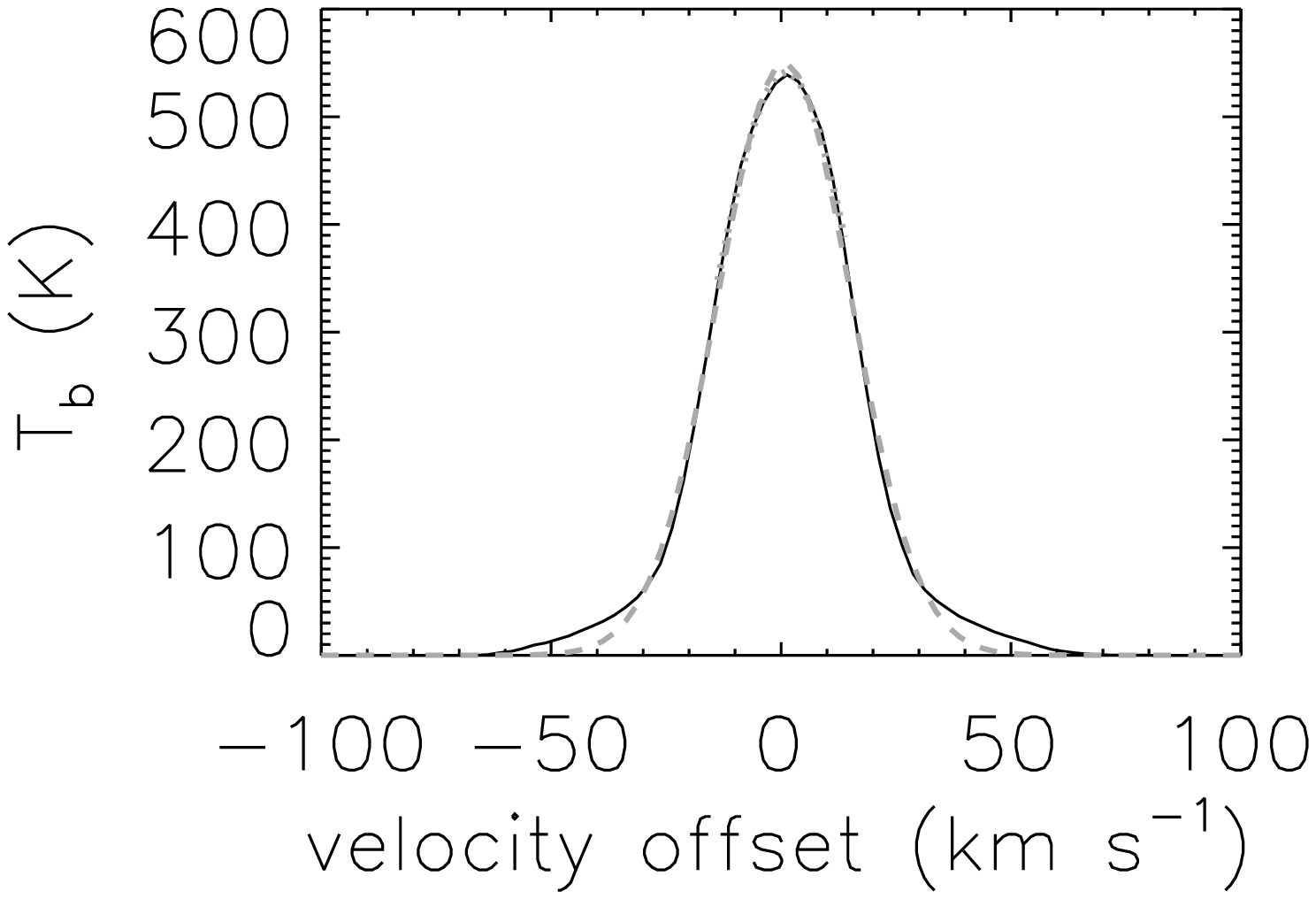}
\caption{\ion{H}{1} line profiles in the four ISM models viewed from the top.  
({\it Top row}) 1x, 8x, ({\it bottom row}) 64x, and 512x models.  
Single Gaussian function fits, shown in grey dashed curves,   
poorly match the line profiles, particularly for the 8x model, 
because of significant broad wings. 
Hence, we additionally fit only the central $\pm 15$ km s$^{-1}$ of the 
line profiles with 
narrower Gaussian functions shown in grey dotted line.  
The \ion{H}{1} linewidths measured near  
the line centers change relatively little as the SN rate goes up.  The tiny 
dip near the emission peak in the 512x model is caused by self-absorption 
of \ion{H}{1} emission. 
\label{h1line}}
\end{figure}

Figure \ref{h1line} displays the 21 cm \ion{H}{1} emission line profiles 
for the four ISM models as viewed face-on.  
With the exception of the 8x model, the line shapes are 
similar to the universal profile found by Petric \& Rupen (2007).  
As shown, the profiles are wider than single Gaussian fits.  
The broad wings may be due to the warm component of the neutral gas. 

We estimate the linewidth, $\sigma_{\rm v,H~I}$, by fitting a single Gaussian 
to each line profile 
(grey dashed curves).  The Gaussian fits poorly match the line profiles in 
all models (although it is particularly bad for the 8x model), 
because of the broad wings, 
so we additionally fit only the central $\pm 15$ km s$^{-1}$ of the 
line profiles with 
narrower Gaussian functions shown in grey dotted lines.  
The broad wings in the 8x model are due to warm gases that are more 
prominent at lower temperatures ($4000 \lesssim T \lesssim 7000$ K) 
than in the other models (see Fig. \ref{vfrac}b).  If we set the maximum 
cutoff temperature for neutral gas to 4000 K instead of 7000 K, 
the wings show substantially decreased flux as in the other models.  
On the other hand, 
removing the gas at high altitudes ($|z| \ge 0.25$ kpc) 
does not affect the line profile, which indicates that the broad wings are not 
caused by a galactic wind.  

We find that the linewidths (fit to the central $\pm 15$ km s$^{-1}$) 
   $\sigma_{\rm v, H~I}$ 
are nearly constant across a wide range of SN rates 
as given in Table~\ref{tbl_results},
similar to the behavior of the mass-weighted velocity dispersions.  
These linewidths are comparable to the velocity dispersions 
found by Agertz et al.\ (2008) 
in global galaxy models.  They attributed the turbulent motions to global 
non-axisymmetric modes and cloud-cloud tidal interactions and merging.  
We find that SN explosions drive turbulent motions that are 
at least comparable in terms of energy. 
The \ion{H}{1} linewidths, associated with only neutral hydrogen, 
are systematically higher than the mass-weighted velocity dispersions.  
Hence, using \ion{H}{1} linewidths to infer the overall velocity dispersion 
may lead to an overestimate. 
Our results are 
consistent with the observed near constancy of line-of-sight velocity 
dispersions with the galactocentric radius (Dickey et al.\ 1990; Kamphuis 1993), 
and extend the numerical results of Dib et al.\ (2006) 
to stratified interstellar media with gas surface densities 
increasing with SN rate.

\subsection{Thermal vs. Turbulent Pressures}

\begin{figure*}
\epsscale{1.0}
\plotone{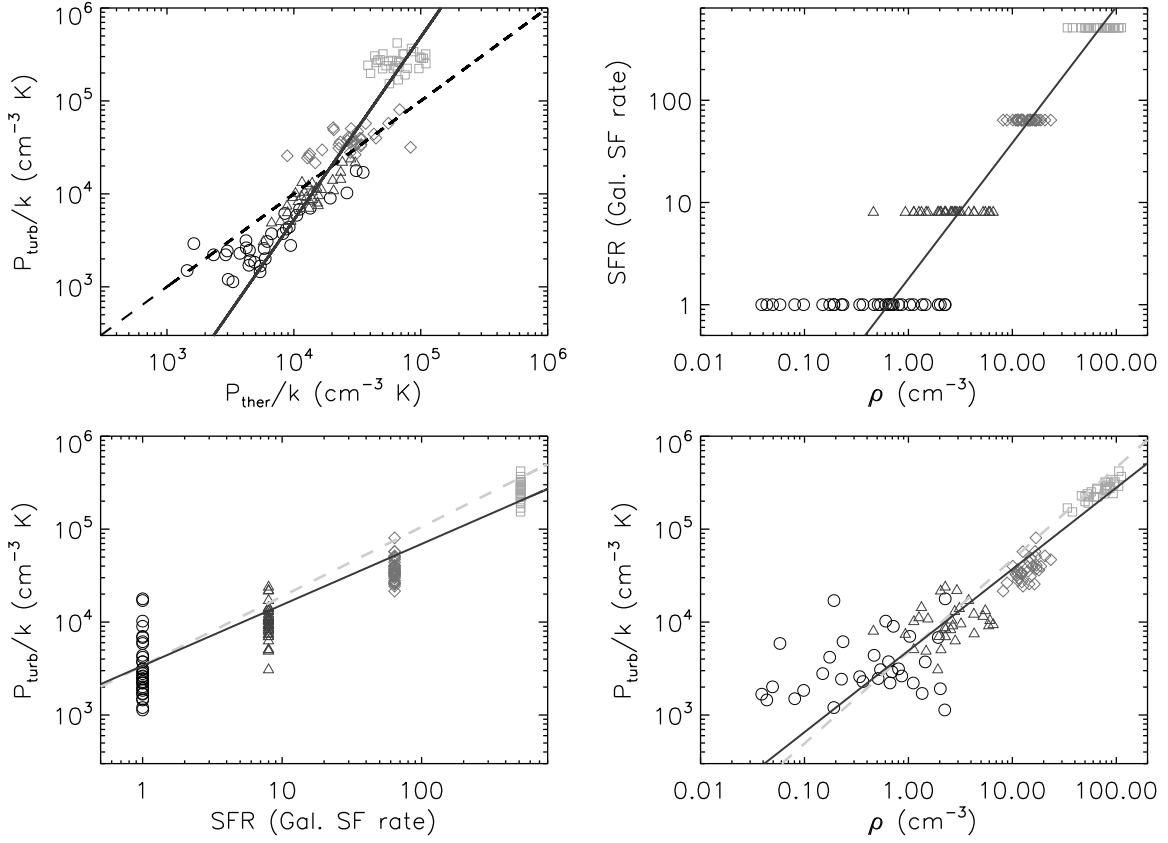}
\caption{Turbulent pressure variation in our four models with different SN rates. 
Displayed are: 
(a) turbulent vs. thermal pressure, 
(b) assumed SFR vs. density, 
(c) turbulent pressure vs. SFR, 
and 
(d) turbulent pressure vs. density. 
Each quantity is evaluated 
in (125 pc)$^3$ boxes near the midplane of the model galaxies. 
Different symbols indicate the ISM models:  
1x (circles), 8x (triangles), 64x (diamonds), and 512x (squares).   
In (a), the black dashed line is drawn for $P_{\rm turb} = P_{\rm ther}$.
The dark solid lines represent best-fit lines 
to all four ISM models.  
The grey dashed lines in (c) and (d) represent the relations expected from 
the subgrid model if the ratio $\epsilon'_1/\epsilon_2$ is constant,  
equations (\ref{ptu_rho}) and (\ref{ptu_edot}), respectively.  
\label{subgrid_plot}}
\end{figure*}

Figure \ref{subgrid_plot}a shows a remarkably 
tight correlation between thermal pressure 
and turbulent pressure (computed in boxes with 125 pc sides) 
across our ISM models.  
This arises because higher SN rates 
naturally lead to increases in both thermal and turbulent pressures. 
The best-fit line has a slope that is steeper than unity 
($1.97 \pm 0.13$), 
implying that 
    the turbulent pressure increasingly dominates 
    at higher SN rates.
Note that this does not necessarily mean that turbulent pressure is 
comparable in magnitude to thermal pressure on all scales.  
Turbulence from large-scale motions (i.e., 
scales larger than our box size that are not well modeled by the simulation) 
may have a significant contribution, 
and so $P_{\rm tot}/P_{\rm ther}$ may be significantly larger than $\sim$2 on 
scales above 125 pc.

\subsection{Kinetic Energy Spectrum}
\label{kespec}

\begin{figure}
\epsscale{1.25}
\plotone{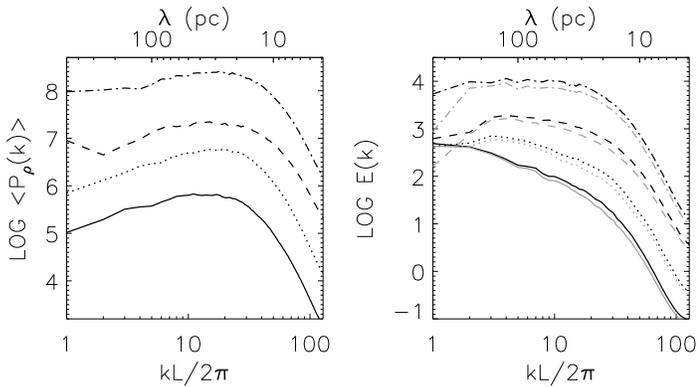}
\caption{{\it Left:} Angle-averaged density power spectra. 
{\it Right:} Kinetic energy spectra ({\it black}) and angle-averaged 
velocity power spectra ({\it grey}).  The kinetic energy is distributed 
over a broad range of wavelengths, but the characteristic scale containing 
most of the energy shifts to larger wavenumbers (shorter wavelengths) 
as the SN rate increases.  
\label{pspec}}
\end{figure}

In Paper I, it was shown that the kinetic energy in an explosion-dominated medium 
is distributed over a broad range of wavelengths.  Then how do the characteristic 
scales of the density and kinetic energy distributions change as a function of SN rate?  
Figure \ref{pspec} displays angle-averaged density power spectra ({\it left}) 
as well as 
kinetic energy spectra in black and angle-averaged 
velocity power spectra in grey ({\it right});  see Paper I for details on 
how they are computed.  
As in the Galactic SN rate model, the velocity power spectra 
deviate significantly from the Kolmogorov spectrum, 
which is applicable to an incompressible 
medium driven at a single large scale:  these assumptions are not 
applicable to the ISM 
driven by supersonic shocks on a range of scales.  
It is clear from Figure \ref{pspec} that the scale containing most of the 
kinetic energy 
shifts to larger wavenumbers (smaller wavelengths) with increasing SN rate. 
We quantify this change by computing the effective driving scale\footnote{Note 
that the term ``driving scale'' is a misnomer since there is no 
single scale where energy is injected, unlike in Kolmogorov's idealized picture 
of incompressible turbulence.} defined by 
\begin{equation}
\label{ldeff}
L_{\rm d,eff} \equiv \frac{2 \pi \int k^{-1} E_k \, dk}{\int E_k \, dk} \; . 
\end{equation}
We find 
    the values of  $L_{\rm d,eff}$ given in Table~\ref{tbl_results}.
The best-fit line has a form 
$L_{\rm d,eff} \approx (126 \pm 18 \; {\rm pc}) \, (\dot{\Sigma}_{*}/\dot{\Sigma}_{*,1})^{\; -0.15 \, \pm \, 0.03}$, 
where $\dot{\Sigma}_{*,1}$ is the SN rate per unit area in the 1x model.
It should be remembered that we are dealing only with SN driven turbulence.  
Large-scale turbulence driven by, e.g., self-gravity may contribute to turbulence 
on larger scales, 
so that the combined power spectrum may be more or less continuous out to 
kiloparsec scales, as observed (e.g., Elmegreen et al.\ 2001).

\section{Subgrid Model for Turbulent Pressure}
\label{subgrid}

\subsection{Basic Features}

A genuine, three-dimensional, cosmological model that properly includes SN feedback 
would need to encompass an unrealistically 
large dynamic range, starting from length scales responsible for galaxy 
formation ($>$1 Mpc) down to those relevant to star formation ($<$1 
pc). The resulting spatial dynamic range of at least $10^6$ 
requires computational resources currently impracticable.
Past attempts to circumvent this problem include analytic 
approaches based on McKee \& Ostriker (1977)'s multiphase model of 
the ISM (e.g., Efstathiou 2000) and semi-analytic methods 
with parametrized prescriptions for star formation and feedback 
(Kauffmann et al.\ 1993; Somerville \& Primack 1999; Cole et al.\ 2000).
 
An alternative method is to use subgrid (or subresolution) models 
(Yepes et al.\ 1997; Springel 2000; Semelin \& Combes 2002; Springel \& Hernquist 2003). 
In these models, one represents 
the physics unresolved in the global model (i.e., turbulent motions 
on small scales) by an analytic prescription. Relevant physical quantities 
are evolved on a subgrid scale, and their appropriate averages 
yield a simple description of the medium on the smoothing scale. The desired 
net effect is to have a large-scale simulation on a coarse grid, 
indistinguishable from the one in which all motions are resolved down 
to parsec scales.

Our approach is inspired by the subgrid model described by 
Springel (2000), which assumes that 
turbulent motions below resolved scales produced by SN feedback can 
be represented by a second reservoir of internal energy of the gas, 
a term that helps to support the gaseous 
medium in the absence of substantial thermal support.  
Nonlinear physical processes such as shock compressions, radiative cooling and 
thermal instability shape 
the properties of interstellar turbulence.  Hence, 
a subgrid model based on heuristic analytic arguments should be checked by 
direct numerical simulations that include the relevant physics. 
For this reason, we explicitly calibrate our subgrid model against 
the results of our local ISM simulations.  
The subgrid model is 
therefore based on the results of our  
high-resolution local ISM models 
that resolve turbulent motions of the gas. 
In a sense, 
we imagine an entire model galaxy on a coarse grid ($\sim$100 
pc spatial resolution), and within it, identify each cell with a 
box with side $\sim$100 pc taken from one of our local models. 

The key parameter in our subgrid model is the {\it turbulent pressure}. 
Local ISM models described in this paper and elsewhere suggest that 
turbulent pressure is more important than either thermal 
(Boulares \& Cox 1990; Lockman \& Gehman 1991) 
or magnetic pressures (Kim 2004; de Avillez \& Breitschwerdt 2005) in most ISM 
conditions. This suggests that cosmological subgrid models should include 
a term for the turbulent pressure, rather than focusing on thermal pressure alone. 
Here we suggest a simple extension where we add an additional energy term 
in the fluid equation to describe the turbulent energy on scales not resolved 
by the simulation (Joung 2006; Bryan 2007). 
Two characteristics of our local ISM models are conducive to developing 
such a subgrid model for turbulent pressure as a way to represent small-scale 
motions. First, $>$90\% of the total kinetic energy is on scales below 
200 pc (Paper I). In current high-resolution cosmological simulations, 
therefore, most of the turbulent energy lies on scales smaller than a resolution 
element, and there is hope that such a model assuming a separation of 
scales would work. Second, the turbulent pressure is nearly constant 
on scales greater than $\sim$100 pc (\S~\ref{pequil}), which suggests that 
for a given SN rate, a single value of turbulent pressure can characterize 
the medium regardless of the local gas density, albeit with the scatter shown 
in Figure \ref{sigs}b ({\it left}). 

In our subgrid model, the pressure is given by 
\begin{equation}
\label{ptot}
P = P_{\rm ther} + P_{\rm turb} = (\gamma-1) \, \rho \, (e+q)
\end{equation}
where $e$ is the specific thermal energy density, while $q = \sigma_{\rm turb}^2$ 
is the specific turbulent energy density. 
Essentially, we require an additional 
energy equation that is similar to the one for thermal energy but includes 
a source term reflecting the turbulent energy input from SNe instead of 
thermal heating terms, and a sink term corresponding to the decay of 
turbulence instead of radiative cooling terms:
\begin{equation}
\label{dqdt}
\frac{dq}{dt} = -(\gamma-1) \, q \bnabla \cdot \bov + \epsilon_1 \frac{\epsilon_{\rm fb} \, \dot{\rho}_*}{\rho} - \epsilon_2 \frac{q^{3/2}}{L} \, . 
\end{equation}
The star formation rate per unit volume, 
$\dot{\rho}_* = c_* (\rho/t_{\rm dyn})$, where $t_{\rm dyn}=[3\pi/(32 G \rho)]^{1/2}$ is 
the dynamical time and the star formation efficiency per 
dynamical time, $c_* \approx 0.02$--$0.03$ (Krumholz \& Tan 2007).
The local energy input due to SNe, 
$\dot{\varepsilon}_{\rm in} = \epsilon_{\rm fb} \dot{\rho}_*$,
where the energy input per unit stellar mass, 
$\epsilon_{\rm fb} = 4 \times 10^{48}$ erg M$_{\odot}^{-1}$ 
if the initial mass function (IMF) with a slope of $-1.5$ between 0.1 and 40 M$_{\odot}$ 
and all stars with mass $>$8 M$_{\odot}$ exploding as SNe.
The turbulent energy is assumed to decay at a rate proportional to $q/t_{\rm cross}$ 
where $t_{\rm cross} = L/q^{1/2}$ is the crossing time at the turbulent driving scale 
(Mac Low 1999); 
we assume $L \approx L_{\rm d,eff}$, calculated in \S~\ref{kespec}. 

The two dimensionless parameters $\epsilon_1$ and $\epsilon_2$ are selected 
to match local ISM simulations.  
The first of these parameters, $\epsilon_1$, measures the efficiency of conversion 
of SN energy into kinetic motions, 
which may depend on, e.g., the density and metallicity of the surrounding medium 
(e.g., Thornton et al.\ 1998).  
We expect $\epsilon_1$ to decrease with increasing SN rate.  
The second parameter, on the other hand, may be more or less constant;  
Mac Low (1999) found 
$\epsilon_2 \simeq 0.84$ for a range of energy injection rates and driving scales 
from a large set of 
hydrodynamic and magnetohydrodynamic simulations with an isothermal equation of state. 

The thermal energy equation, in turn, is given by
\begin{equation}
\label{dedt}
\frac{de}{dt} = -(\gamma-1) \, e \bnabla \cdot \bov + \frac{n \Gamma - n^2 \Lambda}{\rho} + (1-\epsilon_1) \frac{\epsilon_{\rm fb} \, \dot{\rho}_*}{\rho} + \epsilon_2 \frac{q^{3/2}}{L} \, ,
\end{equation}
where $n \Gamma(\rho, e, I_{\rm isrf})$ and $n^2 \Lambda(\rho,e)$ denote 
appropriate thermal heating and radiative cooling rates per unit volume, 
respectively, $n$ is the number density of the gas, 
and $I_{\rm isrf}$ is the intensity of the interstellar radiation field. 
The last two terms account for direct thermal heating from SNe 
and the dissipation of turbulent energy into heat. When combined with the Euler equations, 
equations (\ref{ptot}), (\ref{dqdt}) and (\ref{dedt}) determine the evolution of thermal 
and turbulent energy densities.

\subsection{Calibration with Local ISM Models}

We can make simple analytic arguments to see how turbulent pressure is 
expected to vary as a function of the SN rate. To do this, we assume a 
steady state,
where the energy input rate from SN explosions, $\dot{\varepsilon}_{\rm in}$, is balanced 
by the dissipation of turbulent motions. We set $dq/dt = 0$ and $\bnabla \cdot \bov = 0$ 
in equation (\ref{dqdt}) to obtain
\begin{equation}
\epsilon'_1 (1-f_{\rm w}) \, \dot{\varepsilon}_{\rm in} = \epsilon_2 \, \frac{\rho \, q^{3/2}}{L} \; ,
\end{equation}
where we write $\epsilon_1$ as $\epsilon'_1 (1-f_{\rm w})$ to explicitly account for 
the dependence on $f_{\rm w}$, the fraction of input energy that goes into 
driving large-scale ($\gtrsim 100$ pc) motions such as galactic winds, and hence 
is not captured in the subgrid model. 
Now, to facilitate comparison with our ISM simulations, where 
we used the Schmidt-Kennicutt law for the initial conditions (Eq. \ref{sigmastar}), 
we adopt $\dot{\varepsilon}_{\rm in} 
= \epsilon_{\rm fb} \, A (2 \rho H_{\rm gas})^{1.4}/(2 H_{\rm SN})$, where $H_{\rm gas}$ and $H_{\rm SN}$ are vertical scale 
heights for gas and SNe, respectively. The turbulent pressure is then given by 
\begin{eqnarray}
\label{pturbsg}
P_{\rm turb} &=& \rho \, q 
= \rho \left( \frac{\epsilon'_1 (1-f_{\rm w})}{\epsilon_2} \frac{\dot{\varepsilon}_{\rm in} L}{\rho} \right)^{2/3}\\ 
&=& \, \mathcal{C} \, [ (1-f_{\rm w}) \, H_{\rm gas}^{7/5} \, L ]^{2/3} \, \rho^{\, 19/15} \, ,
\end{eqnarray}
where the combination of parameters, 
\begin{equation}
\mathcal{C} \equiv \left( \frac{2^{0.4} \, \epsilon'_1}{\epsilon_2} \, \frac{\epsilon_{\rm fb} \, A}{H_{\rm SN}} \right)^{2/3} \, .
\end{equation}

Now, consider each 
    factor
in brackets on the RHS of equation (\ref{pturbsg}).  
The driving scale has
already been computed (Eq. \ref{ldeff}); we take 
$\, L \approx L_{\rm d,eff} \propto \dot{\Sigma}_*^{\; -0.15 \, \pm \, 0.03} \propto \dot{\varepsilon}_{\rm in}^{\; -0.15 \, \pm \, 0.03}$, 
if $\epsilon_{\rm fb}$ and $H_{\rm SN}$ change little with the SN rate.  
To compute the gas scale height, $H_{\rm gas}$, 
we plot the SN rate (normalized to the Galactic rate) as a function of gas density 
in Figure \ref{subgrid_plot}b.  
If $H_{\rm gas}$ is constant across our models, 
the energy input rate will scale with the gas density in the same way 
as it varies with the gas surface density 
($\, \dot{\Sigma}_* \propto \rho^{\, 1.4} \,$).  
This is indeed close to what we find for the best-fit line 
shown in Figure \ref{subgrid_plot}b: 
   \begin{equation}    
     \dot{\Sigma}_{*} = (1.79 \pm 0.17) \, \dot{\Sigma}_{*,1} \, \left( \frac{\rho}{\rho_1} \right)^{\,
       1.33 \, \pm \, 0.03} \; ,
    \end{equation} 
where $\rho_1 \simeq 0.7$ cm$^{-3}$ is the mean density in the midplane of the 1x model.  
From this relation and $H_{\rm SN} \approx 90$ pc, 
we infer a nearly constant gas scale height of 
    170 pc $(\dot{\Sigma}_{*}/\dot{\Sigma}_{*,1})^{\, 0.04 \, \pm \, 0.02}$.  

To estimate how the fraction of energy 
driving large scale motions ($f_{\rm w}$) scales with the SN rate, 
we distinguish between the small-scale, turbulent motions within 
boxes having 125 pc sides, as examined in \S~\ref{turbpres}, 
and the bulk motion of these individual boxes.  
This amounts to comparing the sum of the kinetic energy contained in 
(mass-weighted) velocity dispersions inside the boxes ($\rho \, \sigma_{\rm turb}^2/2$) 
and that from the 
center-of-mass velocities of the boxes themselves ($\rho | \bov_0 |^2 /2$).  
Using this procedure, 
we obtain 
the values of $f_{\rm w}$ given in Table~\ref{tbl_results}.
This does not contradict our earlier result that almost all the turbulent energy is 
contained on scales below $\sim$200 pc, because we were then discussing 
the kinetic energy spectrum in the midplane region only.  In fact, if we confine 
    our attention to
the region
    with
$|z| \le 125$ pc and compute the energy fraction 
in large-scale motions, we obtain significantly smaller values 
(0.084, 0.065, 0.149, and 0.455).  
In other words, most of the kinetic energy on large scales comes from the bulk 
motions at high altitudes:
   galactic winds.
If we fit a straight line through 
    the values of $f_w$ given in Table~\ref{tbl_results}, we find
\begin{equation}
1-f_{\rm w} = (0.65 \pm 0.15) \, \left( \frac{\dot{\Sigma}_{*}}{\dot{\Sigma}_{*,1}} \right)^{\; -0.06 \, \pm
  \, 0.05},
\end{equation}
   which is effectively a constant.

To study the variation of turbulent pressure in our ISM simulations, 
we compute $P_{\rm turb}$ in 
boxes of size (125 pc)$^3$ taken from the midplane region ($|z| \le 125$ pc) 
of the four ISM models. 
The results
   given in 
Figure \ref{subgrid_plot}c--d show the turbulent pressure 
depends on the input SN rate as well as the gas density averaged over 
cubes with 125 pc sides. 
The expectations from the subgrid model 
    (Eq.\ 
\ref{pturbsg}) 
assuming constant $\mathcal{C}$ 
are shown as grey dashed lines in Figures \ref{subgrid_plot}c 
and \ref{subgrid_plot}d.  
The best-fit lines to the data, shown 
in dark solid lines, 
have the form
\begin{equation}
\label{ptu_edot}
P_{\rm turb} \propto \dot{\Sigma}_*^{\; 0.66 \, \pm \, 0.04} \; .
\end{equation}
and, in terms of the gas density, 
\begin{equation}
\label{ptu_rho}
P_{\rm turb} \propto \rho^
{0.88 \pm 0.05} \; ,
\end{equation}
   This
effective equation of state (EOS) is 
slightly more compressible than
an isothermal EOS, for which $P \propto \rho^{\, 1}$. 
In contrast, 
    much stiffer 
effective EOSs were adopted in previous 
subgrid models.  For example, Springel (2000) used an EOS of the form 
$P \propto \rho^{\, 2}$, obtained by assuming that all SN energy 
first goes into turbulent energy and then converts to thermal energy with 
an efficiency that depends specifically on gas density 
($\epsilon_1 = 1$ and $\epsilon_2 \propto \rho^{\, -1/2}$ in our notation).  
In the hybrid multiphase model developed by Springel \& Hernquist (2003), 
hot gas from SNe 
delays radiative cooling and provides the 
necessary additional pressure.  Here, 
the effective EOS was also 
   stiffer
than isothermal above a threshold density ($\sim$0.1 cm$^{-3}$), 
where star formation was assumed to switch on.  
Our result indicates that SN-driven turbulence does not naturally provide 
a stiff EOS on 100--200 pc scales
but rather that the effective EOS of the ISM averaged over these 
scales 
is slightly sub-isothermal 
(see \S~\ref{discuss} for further discussion).  

The simulation results deviate slightly from the naive 
expectations based on constant $\mathcal{C}$, especially in the models for 
starbursts (the 64x and 512x models).  This 
suggests that $\mathcal{C}$ 
may 
be close to but not exactly a 
constant.  
Combining the scaling relations we have found, we obtain 
\begin{equation}
\mathcal{C} \propto \dot{\varepsilon}_{\rm in}^{\; -0.55 \, \pm \, 0.07} \; ,
\end{equation}
or $\epsilon'_1 \propto \dot{\varepsilon}_{\rm in}^{\, -0.82 \, \pm \, 0.11} \propto \rho^{\, -1.15 \, \pm \, 0.15}$ 
for a constant $\epsilon_2$.  
Hence, a smaller fraction of SN energy is used to drive kinetic motions 
as the SN rate (and the mean gas density) increases, 
although the dependence is weak.

We must mention several caveats that may affect our conclusion.  
Our ISM simulations do not include magnetic fields or self-gravity of gas, 
which can change the structure and the pressure distribution of the medium, 
especially in the dense regions.  
In principle, unphysical cooling of numerical origin
   may be important,
particularly in the 64x and the 512x models.  This
    would occur at the interfaces between hot and cold gas, which are
    inevitably wider than the physical value, and so contain more gas
    at intermediate temperatures between $10^4$~K and $10^6$~K that is
    subject to strong radiative cooling than they physically
    should (e.g., Mac Low et al.\ 1989). Such cooling could yield values of
$P_{\rm turb}$ that are lower than the correct values.  
However, the resolution study discussed in \S~\ref{sigma} 
    shows that 
   this is a surprisingly minor problem in practice.
An alternative method to examine this issue 
    would be
to use a tracer field
    (Yabe \& Xiao 1993) 
to suppress the excessive cooling in young SN remnants as in previous work 
(Mac Low \& Ferrara 1999; Fujita et al.\ 2004), 
after appropriately modifying the criterion for suppression for the case of 
multiple discrete SN explosions. 

\section{Discussion}
\label{discuss}


\paragraph{Comparison with Chandrasekhar's formula for the effective sound speed.} 
In \S~\ref{turbpres}, we have used 
$P_{\rm tot} = \rho \, (\sigma_{\rm ther}^2+\sigma_{\rm turb}^2) = \rho \, (\sigma_{\rm ther,b}^2/\gamma+\sigma_{\rm turb,b}^2/3)$. 
The reader may wonder 
why this is the appropriate expression. For example, compare it with the expression 
$c_{\rm s,eff}^2 = c_{\rm s}^2 + \langle u^2 \rangle /3$ where $c_{\rm s,eff}$ is the effective 
sound speed, which Chandrasekhar (1951) used to extend Jeans's analysis to  
an infinite homogeneous turbulent medium. One immediately notices that, in his 
formulation, $c_{\rm s}^2$ 
was used instead of $c_{\rm s}^2/\gamma$. To understand this, we propose the following 
thought experiment. If we have two parcels of gas at the same temperature but with two 
different adiabatic indices (e.g., $\gamma$ = 5/3 and 7/5), they should have the same 
pressure (since $P_{\rm ther} = nkT$). However, when you try to compress the gases, they have different 
resistance to the compression, i.e., the one with higher $\gamma$ is 
more difficult 
to compress and 
hence their sound speeds differ. The first case is relevant to pressure equilibrium in our models; 
the second to Chandrasekhar's problem. Essentially, our problem is a pressure issue, 
while Chandrasekhar's is a communication timing issue (the Jeans stability criterion 
says $t_{\rm cross} < t_{\rm dyn}$).


\paragraph{Large-scale turbulence driving.} 
We argued that most of the kinetic energy due to SN-driven turbulence can be 
contained in resolution elements (with size of 100--200 pc) of cosmological simulations, hence 
allowing the approach of subgrid modeling. In reality, however, there are 
numerous other energetic processes that occur on larger scales that may 
affect the dynamics. An example would be gravitational instability induced 
by spiral structure such as 
the magneto-Jeans instability (Kim \& Ostriker 2002). 
Li et al.\ (2005a, 2006) argued, using an isothermal equation of state 
to represent stellar feedback, that gravity controls when and where 
star formation occurs, and reproduced the Schmidt-Kennicutt law 
with realistic star formation thresholds. The use of a subgrid 
model for interstellar turbulence does not mean that we ignore these larger-scale, 
possibly gravity-induced motions. Rather, we are making an assumption that 
the characteristic lengths at which these two processes operate are 
sufficiently far apart that they can be treated separately in global models.  
One way to test this hypothesis is to compare simulated and observed column 
density power spectra (Elmegreen et al.\ 2001) for both local and 
global simulations of galaxies. 


  \paragraph{Observed linewidths in ULIRGs.} 
High velocity dispersions in molecular gas, 
$\sigma_{\rm CO} \approx 100$ km s$^{-1}$, were observed 
in ultraluminous infrared galaxies (ULIRGs) such as Arp 220 (Downes \& Solomon 1998). 
Murray et al.\ (2005; see also Thompson et al.\ 2005) suggested that in extreme 
starbursting systems such as ULIRGs, the ambient high density gas 
impedes the expansion of SN remnants, drastically lowering the 
efficiency with which SN energy escapes the host galaxy.  
Their model does not include the effect of superbubbles: 
correlated SNe with high $n_{\rm SN}$ may decrease the ambient density near 
explosion sites, and the overall radiative cooling rate in such explosions 
may be significantly reduced. 
    In our models with high SN rates, SN explosions near the midplane drive 
    large-scale motions about as effectively as in models with lower SN rates.  
    The fraction of energy stored in turbulent motions as opposed to heat 
    increases with SN rate.
We note that 
Fujita et al.\ (2008) recently found that extremely superthermal linewidths 
($\sim$320 km s$^{-1}$) in 
    \ion{Na}{1}
absorbing gas could be reproduced 
using a model for an energy-driven superbubble 
with a spatial resolution of $\le 0.2$ pc, 
assuming a single central starburst region and axisymmetry. 
The energy input rate of our 512x model is 
$\sim$1.3$\times$10$^{41}$ erg s$^{-1}$, comparable
to the lowest mechanical luminosity model in Fujita et al.\ (2008).  

\paragraph{Effective equation of state.} 
The models presented in this paper 
    demonstrate
that a stiff EOS does not result from SN explosions in 
    a
stratified ISM, since 
the effective EOS including turbulent pressure across the local ISM models 
is close to isothermal. 
   In the past, stiff EOSs were proposed as a way of preventing the
   collapse of disk galaxies to unphysically small radii in
   cosmological simulations.  However, Li et al.\ (2005b) showed that
   such collapse was due to violation of the Jeans criterion (Truelove
   et al. 1997) in low resolution simulations, and that fully
   resolved, isothermal disks do not collapse.
   Simulations of full galaxies including our sub-grid model will
   reveal whether there are further effects that a stiffer equation of
   state would better reproduce.

\paragraph{Limitation of the steady state assumption.} 
The subgrid model developed in this paper can be easily implemented in 
cosmological simulations.  
However, our approach is 
limited by the assumption of constant SN rate; we dealt only with steady state systems. 
The model is appropriate if the characteristic timescale $\tau_{\rm SF}$ 
over which the SN (SF) rate averaged over $\sim$100 pc regions changes 
appreciably is much longer than the time it takes for the system to reach a 
steady state ($\sim$60 Myr in our simulations). The assumption is valid, e.g., 
in merging galaxies, where $\tau_{\rm SF}$ is on the order of 0.1--1 Gyr.

\section{Summary}
\label{summary}

Interstellar turbulence is thought to play a major role in the 
formation and support of molecular clouds and the regulation of the size, thickness, 
and star formation rate of galactic disks (Mac Low \& Klessen 2004; 
Elmegreen \& Scalo 2004). We examined 
the physical characteristics of the ISM by constructing high-resolution 3D models 
of a 
stratified ISM driven by both correlated and isolated SN explosions. 
A suite of numerical experiments, each with a simulated 
volume of (0.5 kpc)$^2$$\times$(10 kpc), were performed using an adaptive 
mesh code at a maximum spatial resolution of 1.95 pc; they included vertical gravity 
and parameterized heating and cooling. The simulations cover wide ranges of supernova 
rates and vertical gas column densities, with initial conditions 
based on the Schmidt-Kennicutt law (Kennicutt et al.\ 1998, 2007). 

We find that the turbulent velocity dispersion is inversely 
proportional to the square root of the local density: 
$\sigma_{\rm turb} \propto \rho^{-1/2}$ 
on $\gtrsim$50 pc scales. 
The turbulent pressure, $\rho \, \sigma_{\rm turb}^2$, in such a medium is 
nearly constant 
at a given scale, 
even though the gas density varies by about seven orders of 
magnitude. This suggests that, for a given SN rate, a single 
value of turbulent pressure can characterize the medium regardless 
of the local gas density. When combined with the finding in Paper I that 
$>$90\% of the total kinetic energy is contained on scales 
below 200 pc, the two 
charateristics of the ISM are conducive to developing a subgrid model 
for turbulent pressure as a way to represent small-scale motions. 
One such model was developed in this paper and explicitly calibrated using 
the local ISM simulations.  

We find that 
the mass-weighted velocity dispersion $\sigma_{\rm turb}$
    and the simulated \ion{H}{1} linewidth $\sigma_{\rm v, H~I}$ are
nearly constant across 
a range of 
512 in SN 
    rate (see Table~\ref{tbl_results} for values).  
In other words, 
the appropriate equation of state for $P_{\rm turb}$ in gas averaged over $\sim$100 
pc regions is close to isothermal. 
    In our highest supernova rate model, superbubble blow-outs occur, and 
    the turbulent pressure on large scales is $\gtrsim$4 times higher than 
    the thermal pressure.

   We propose a subgrid model that
naturally includes the effect of turbulence in large-scale 
simulations, at least as far as we understand it from
   our 
small-scale models, and furnishes a sound, physically 
motivated prescription for including the effect of SN feedback in cosmological 
simulations.  It treats 
turbulence as strictly a pressure term, although small-scale effects on the 
star formation rate  
could, in principle, be included in a statistical way. The advantage of this 
type of approach 
is that we prescribe the subgrid model from what we know (bottom-up) rather than 
what we need in order to resolve current problems in cosmological simulations (top-down).

\acknowledgments

We are grateful to I. Goldman, Y. Li, J. Oishi, J. Stone, 
E. V\'{a}zquez-Semadeni, and T. Thompson for stimulating discussions.
We thank R. Cen and B. Draine for useful discussions on  
the scaling of 
    the
diffuse heating rate and 
J. Maron for proposing the thought experiment described in \S~\ref{discuss}.
The software 
used in this work was in part developed by the DOE-supported ASCI/Alliance 
Center for Astrophysical Thermonuclear Flashes at the University of Chicago.
Computations were performed at the Pittsburgh Supercomputing Center and at the 
National Center for Supercomputing Applications supported 
by the NSF.

\end{document}